\documentclass[aps,twocolumn,groupedaddress,showpacs,showkeys]{revtex4}

\usepackage{amssymb,amsmath,amsbsy} 
\usepackage{graphicx}
\usepackage{units}
\usepackage{verbatim}
\usepackage{xcolor}

\def\ifcomment{\iffalse}
\def\changed{\textcolor{black}} %{purple}}  
\def\added{\textcolor{black}} %{blue}}  

\def\pdf{f}
\def\Ham{H}
\def\Mom{P}
\def\Hconstraint{\CH}
\def\Hhat{\hat \Hconstraint}
\def\Phat{\hat  {\underline  P} }
\def\Vel{v}

\def\time{t} 
\def\coord{x}
\def\pdf{f}
\def\amp{\psi}
\def\phase{\varphi}

\def\Pconstraint{P}
\def\Phat{\hat\Pconstraint}

\def\Ncoord{N}
\def\Jac{\CJ}
 
\def\Amat{\mathsf{A}}
\def\coordOp{\hat x} 
\def\Qhat{\hat Q}

\def\Nb{n}
\def\Nparticles{m}
\def\xDim{d}
\def\Ndof{N}
\def\HamOp{\hat \Ham} 

\def\eqdef{:=}

\def\VelOp{\hat \Vel}
\def\angle{\vartheta}
\def\action{J}

\renewcommand\Im{{\rm Re}\,}
\renewcommand\Im{{\rm Im}\,}
\newcommand\tr{{\rm tr}\,}
\renewcommand\det{{\rm det}\,}
\newcommand\setlist[1]{\left( #1 \right)}
\newcommand\PoissonBracket[2]{\left\{ #1, #2  \right\}}
\newcommand\Commutator[1]{\left[ #1 \right]}
\newcommand{\norm}[1]{{\left| {#1}\right| }}
\newcommand{\abs}[1]{{\left| {#1}\right| }}
\newcommand{\avg}[1]{{\left< {#1}\right> }}
\newcommand\ket[1]{\left|#1\right>}
\newcommand\bra[1]{\left<#1\right|}
\def\Observable{O}

\def\PhaseAction{W}
\def\Lagrangian{L}
\def\Xmax{X}

\def\hPlanck{h}

\def\qcoord{q}
\def\pcoord{p}

\def\Pcoord{P}

\def\charge{q}
\def\mass{m}
\def\vel{v}
\def\VAfield{{\bf A}}
\def\Xmax{X_{\rm max}}

\def\VXhat{{\bf \hat X}}

\def\Vd{{\bf d}}
\def\PoincareForm{ \boldsymbol{\alpha}}
\def\LagrangeForm{ \boldsymbol{ \Omega}}
\def\LagrangeTensor{  \Omega}
\def\PoissonForm{{\mathbf J}}
\def\PoissonTensor{J}
\def\Coord{X}
\def\Coordhat{{\hat X}}
\def\velhat{{\hat\vel}}

\def\Vx{{\bf x}}
\def\Vxdot{{\dot{\bf x}}}
\def\Vxhat{{\bf \hat x}}
\def\Vvel{{\bf v}}
\def\Vvhat{{\bf \hat v}}
\def\Vnabla{{\bf \nabla}}
\def\Vpartial{{\bf \partial}}
\def\VP{{\bf P}}
\def\VPhat{{\bf \hat P}}

\def\Ngrid{L}
\def\Ndim{D}
\def\sparsity{s}
\def\Nparticles{M}

\def\Reals{{\mathbb R}}
\def\Complex{{\mathbb C}}
 \def\EvolutionOp{{\hat \CU}}
 \def\Vxi{{\boldsymbol \xi}}
 \def\nqubit{n}
\def\action{J}
\def\angle{\theta}

\def\ycoord{y}
\def\Vy{{\bf \ycoord}}
\def\VolumeForm{{\boldsymbol \nu}} %{{d\boldsymbol \mu}}
\def\ProbabilityForm{{ \boldsymbol \mu}}%{{d\boldsymbol P}}
\def\Vcoord{V} 

\def\PDF{F}
\def\Amp{\Psi}
\def\Valpha{{\boldsymbol\alpha}}

\def\VLhat{\boldsymbol{ \hat \Pi}}
\def\VZhat{\mathbf{ \hat Z}}
\def\VZhat{\mathbf{ \hat Z}}

\def\GenFunc{S}

\def\Nhat{{\hat N}}
\def\Anglehat{{\hat \Theta}}
 
\def\AsymptoticallyBoundedAboveBy{O}

\def\CH{{\cal H}}

\def\CJ{{\cal J}}
\def\CK{{\cal K}}
\def\CL{{\cal L}}
\def\CM{{\cal M}}

\def\CQ{{\cal Q}}

\def\CS{{\cal S}}

\def\CU{{\cal U}}
\def\CV{{\cal V}}

\begin{document}

%\preprint{LLNL-JRNL-408174}

\title{%The Koopman-von Neumann Approach to 
Koopman-von Neumann Approach to \\
Quantum Simulation of Nonlinear Classical Dynamics} 


\author{Ilon Joseph}
\email[]{joseph5@llnl.gov}
%\thanks{}
%\author{Dmitri D. Ryutov}
\affiliation{Lawrence Livermore National Laboratory}


%\date{February 9, 2010} %first draft
\date{\today}
 
\begin{abstract}
Quantum computers can be used to simulate nonlinear non-Hamiltonian classical dynamics on phase space by using the generalized Koopman-von Neumann formulation of classical mechanics.  The Koopman-von Neumann formulation implies that the conservation of the probability distribution function on phase space, as expressed by the Liouville equation, can be recast as an equivalent Schr\"odinger equation  on Hilbert space  with a Hermitian Hamiltonian operator and a unitary propagator. 
This Schr\"odinger equation is linear in the momenta because it derives from a  constrained Hamiltonian system with twice the classical phase space dimension. 
A quantum computer with finite resources can be used to simulate a finite-dimensional approximation of this unitary evolution operator. 
Quantum simulation of classical dynamics is exponentially more efficient than a \changed{deterministic} Eulerian discretization of the Liouville equation if the Koopman-von Neumann Hamiltonian is sparse.  
\changed{
Utilizing quantum walk techniques for state preparation and amplitude estimation for the calculation of observables leads to a quadratic improvement
over classical probabilistic Monte Carlo algorithms.}
\end{abstract} 
\pacs{03.67.Ac, 03.65.Sq, 05.45.--a}
\keywords{ classical mechanics, quantum mechanics, semiclassical mechanics, Koopman-von Neumann classical mechanics, Koopman operator, Frobenius-Perron operator, quantum-classical correspondence, quantum computing, quantum algorithms, quantum simulation, Hamiltonian simulation}
\maketitle

%\subsection{}
%\subsubsection{}

% If in two-column mode, this environment will change to single-column
% format so that long equations can be displayed. Use
%\begin{widetext}
%\end{widetext}

%Table of Contents
%\tableofcontents

%\emph{1. Introduction.---}
\section{Introduction}
\subsection{Motivation }
In principle, future error-corrected quantum  computers have the power to simulate quantum mechanical systems exponentially more efficiently \cite{Feynman82ijtp,Lloyd96sci} than computers that are bound to satisfy the laws of classical physics. 
The recent growth in the capabilities of today's quantum computing devices has spurred great interest in quantum simulation and they have been used to perform key demonstrations of quantum calculations \cite{Martinez16nat,Roushan17sci,Arute19nat}.
Yet, for many fields of science and engineering, including biology, chemistry, and physics, a large share of today's computational resources are used for the simulation of classical dynamics. Hence, it is important to understand whether quantum computers can provide similar gains in efficiency for the simulation of classical dynamics. 
%The recent U.S. Department of Energy Office of Fusion Energy Sciences report of the 2018 ``Roundtable  on Quantum Information Science'' highlighted the need to understand these issues \cite{SchenkelReport}. 

Classical dynamical systems are typically nonlinear and many important examples are not Hamiltonian. In fact, they are often dissipative.  Since quantum computers can only perform linear unitary operations, it is not clear how nonlinear nonunitary simulations can be performed efficiently. While efficient quantum algorithms for linear ordinary differential equations (ODEs) are known \cite{Berry17arxiv,Costa19arxiv},  an attempt to simulate nonlinear dynamics by measuring the full state at each time step and feeding this information into the next time step would require an exponential amount of resources. The method of Ref. \cite{Leyton08arxiv} is logarithmic in the dimension of the system, but requires an exponential amount of resources in the number of time steps and the polynomial degree of the nonlinearity. 

\subsection{Comparison of Classical and Quantum Resource Requirements}
Often one simulates a finite number of trajectories of a classical dynamical system in order to infer statistics of the evolution of the probability distribution function (PDF), $\pdf$.
A probabilistic description of a classical system is actually similar in complexity to that of a quantum system. If there are $\Nb$ classical bits, then describing the classical PDF, 
\begin{align}
	\pdf=\sum_j \pdf_j \ket{j} \bra{j} \in \Reals^\Ndof,
\end{align}
 over all $\Ndof=2^\Nb$ possible states $\left|j\right>$ requires the specification of  $\Ndof-1$ real numbers: the $\pdf_j$ subject to the constraint $\sum_j\pdf_j=1$. The wavefunction for a pure quantum state holds twice as much information due to the fact that the probability amplitudes are complex. Specifying the wavefunction, 
 \begin{align}
	\amp=\sum_j\amp_j \ket{j} \in \Complex ^\Ndof,
\end{align}
 requires the specification of $\Ndof-1$ complex numbers: the $\amp_j$ subject to the constraint $\sum_j\norm{\psi_j}^2=1$ and, since the overall phase does not matter, a constraint on the phase, such as $ \Im{(\psi_0)}=0$; i.e. $2(\Ndof-1)$ real numbers. Hence,  storing the memory and performing an operation on each component of a probabilistic classical system requires half of the resources of that of a pure quantum state.   
 
 Specifying a mixed quantum state via the density matrix
  \begin{align}
  \rho=\rho^\dagger=\sum_{jk} \rho_{jk} \ket{j}\bra{k} \in \Complex^{\Ndof^2},
  \end{align}
requires specifying $\Ndof^2-1$ real numbers: the $\rho_{jk}$ subject to the constraints of Hermiticity, $\rho=\rho^\dagger$, and unit trace, $\tr{\rho}=1$ (and $\rho$ must also be positive semi-definite).  
The diagonal entries of $\rho$ are sufficient to describe $\pdf$, which implies that a quantum simulation that experiences decoherence by the end of the calculation could still potentially generate a useful simulation of the PDF. On the other hand, a  classical Hamiltonian system is defined on twice the phase space dimension of the quantized version of the classical system, because it  includes both configuration space coordinates and conjugate momentum (or velocity) coordinates, in which case, representing the classical system   also requires $\Ndof^2-1$ real numbers. 
Thus, the resource requirements for probabilistic classical systems and quantum systems are quite similar.

A clear example of this fact is that simulating the Schr\"odinger equation  requires twice as many resources to store the complex amplitude as simulating the diffusion equation for a real density.  Hence, a ``Wick rotation''  is often employed to convert between the Schr\"odinger equation and the diffusion equation. For example, this Wick rotation is used to convert between the temporal propagator of quantum mechanics and the thermal partition function of statistical mechanics \cite{KleinertBook}, which is closely related to the propagator of the diffusion equation. It is also used by a number of quantum Monte Carlo algorithms \cite{KalosMCbook} that seek to find eigenstates of the Schr\"odinger equation by simulating the diffusion equation.

As another example, consider simulation of the Liouville equation, also known as the collisionless Boltzmann or Vlasov equation, which describes the conservation of the classical probability distribution function on phase space and provides the foundation for nonequilibrium statistical mechanics. Such simulations  are commonly performed in diverse fields of science such as population biology, condensed matter physics, molecular dynamics, gravitation, and plasma physics.
 If there are $\Nparticles$ particles traveling in $\xDim$ dimensions, then there are $\Nparticles\xDim$ classical degrees of freedom, and, for a Hamiltonian system, the Liouville equation is a  $\Ndim=2\Nparticles\xDim$ dimensional partial differential equation (PDE). If one employs an Eulerian discretization of phase space, by using $\Ngrid$ spatial grid points in each dimension, then this requires an exponential amount of memory and computational work, $\sim \Ngrid^{\Ndim}$, to process the data. 
 
 Notably, Ref.~\cite{Engel19arxiv} devised a quantum algorithm for solving the linearized Vlasov-Maxwell system that claims to allow the simulation of Landau damping with exponentially reduced computational work.

\subsection{Quantization of Classical Dynamics}  
 There are different strategies that one might potentially employ to represent classical dynamics via quantum mechanics.  Perhaps the most obvious method is to use any valid quantization that reduces to the classical system in the limit of vanishing Planck's constant. The quantum dynamics will track the semiclassical dynamics for arbitrarily long times in the limit that $\hbar\rightarrow 0$. It is known that simulating the quantized version of a classical system can be used to compute dynamical quantities, such as the diffusion coefficient and Lyapunov exponents, more efficiently \cite{Benenti01prl,Benenti05book}. This approach has the benefit that the Schr\"odinger PDE is defined on 1/2 of the classical phase space dimension, and, hence, is technically cheaper to simulate than the Liouville PDE. It also shares similarities  with the manner in which quantum walks can be used to accelerate classical Monte Carlo methods in order to generically achieve a quadratic speedup   
\changed{
  \cite{Aharonov02arxiv, Szegedy04focs, Montanaro15prsa}. 
}
   
 However, for finite $\hbar$, it is known that there are important differences between the classical and quantum behavior \cite{Casati79book,Izrailev80tmf,Chirikov81ssr}. 
 The wavefunction is not as localized as the classical PDF and, due to tunnelling, can spread into regions that are classically forbidden. Thus, the quantized version provides a natural coarse-graining of classical phase space into units of Planck's constant, $\hPlanck$, often included as a normalization constant for entropy in classical statistical mechanics. 
Moreover, different components of the wavefunction carry different complex phase factors which lead to interference effects that are not present in the classical setting.  
Yet another important qualitative difference is the emergence of dynamical Anderson localization \cite{Prange82prl,Chirikov88pd} and many-body localization \cite{Abanin17ap,Alet18crp,Abanin19rmp},  
 which prevents the wavefunction from sampling the entire classically chaotic region accessible to the classical trajectories.  
Finally, while this approach is straightforward  for Hamiltonian systems, it  is not trivial to determine a natural quantization procedure for general non-Hamiltonian dynamical systems.

\subsection{Koopman-von Neumann Approach}  
Nonlinear classical phase space dynamics can be faithfully embedded within a quantum mechanical system -- even for equations of motion that are not Hamiltonian.  
Just after the birth of quantum mechanics, Koopman \cite{Koopman31pnas} and von Neumann \cite{vonNeumann32am1,vonNeumann32am2} realized that classical mechanics can be formulated on Hilbert space in a manner that is exactly analogous to quantum mechanics. The classical Liouville equation, which expresses the conservation of  probability on phase space, and its space-time adjoint, which expresses the evolution of a conserved quantity, can be recast as an equivalent Koopman-von Neumann (KvN)  Schr\"odinger equation \cite{Gorban83preprint,Chirikov88pd}. 
The KvN Hamiltonian (Eq.~\ref{eq:KvN_Hamiltonian})  associated with the KvN equation (Eq.~\ref{eq:KvN_equation})   is linear in the momentum and results from the quantization of an associated constrained Hamiltonian system \cite{Chruscinski03jmp,Chruscinski04jmp} on twice the classical phase space dimension, where the canonical momenta represent Lagrange multipliers that enforce the classical equations of motions as constraints  \cite{Pontryagin62book}. 
 Heisenberg's uncertainty principle applies to each of the original variables and its conjugate momentum, i.e. the Lagrange multiplier, but does not affect any pair of the original variables. Thus, there is complete fidelity to the classical phase space evolution.  

Following the KvN approach in Sec.~\ref{sec:Derivation} leads to the simple derivation of the generalized KvN equation (Eq.~\ref{eq:KvN_equation}), which applies to arbitrary classical dynamical systems, starting from the assumption that the probability distribution, $\pdf$, is the inner product of a complex probability amplitude $\psi$ with its adjoint $\psi^\dagger$, (Eq.~\ref{eq:pdf=|amp|^2}).

To this author's knowledge,  Chirikov, Izrailev, and Shepelyanski  \cite{Chirikov88pd} were the first to publish the generalized form of the KvN equation, which applies to arbitrary classical dynamical systems, which they attribute to an earlier preprint by Gorban and Okhonin  \cite{Gorban83preprint}.  These authors clearly appreciated the meaning of the KvN equation as a first order Hermitian PDE that exactly preserves the original phase space dynamics of the underlying classical system, and, as one which would display the features of classical chaos rather than of quantum chaos.  

Chru\'sci\'nski \cite{Chruscinski03jmp,Chruscinski04jmp} referred to the KvN Hamiltonian as the ``Quantum mechanics of damped systems,'' because a damped system can be embedded within the constrained Hamiltonian discussed in Sec.~\ref{sec:Constrained_Hamiltonian}. While this is a valid quantization of the constrained Hamiltonian, it does not have the physical meaning of being a ``damped'' quantum mechanical system, which would typically be treated using the Master Equation. Rather, it is the KvN Hamiltonian for an arbitrary classical system of ODEs whether damped or not. 

Understanding the properties of the unitary KvN evolution operator, i.e. propagator or transfer operator, corresponding to the Hermitian KvN Hamiltonian should be quite interesting from the point of view of of dynamical systems theory. The closely related Perron-Frobenius  operator and Koopman operator  have been studied and used extensively to characterize invariant measures  \cite{CvitanovicChaosBook} and to develop reduced order models \cite{Mezic04physd, Mezic05nld} for nonlinear dynamical systems.   For divergence-free, i.e. measure-preserving, flows, these three operators are equivalent.  The recognition that the KvN Hamiltonian for dissipative systems still leads to unitary evolution has already found use in the development of high-order conservative numerical discretizations of the advection operator for fluid dynamics \cite{Morinishi98jcp, Morinishi10jcp} and plasma physics \cite{Halpern18pop}.

 The KvN approach can be applied to any system of differential equations that are explicitly presented as first order in time. For example, the method of lines can be used to generate a numerical discretization of a PDE as a finite set of ODEs that can be treated in an equivalent fashion. Hence, the KvN approach can also be applied to important PDEs in classical physics such  as  the Maxwell's equations in conducting media,  the Navier-Stokes equation,  the collisional evolution of the kinetic PDF, and general $N$-body problems in gravitation, plasma physics, and molecular dynamics. 
 Although the probabilistic description of a PDE might appear to be expensive, this is the necessary setting for understanding the evolution of the PDF over regions of phase space and provides the essential framework for uncertainty quantification. It is also necessary for describing interactions with random processes and stochastic forcing terms which are often used to model effects due ``noise'' and  turbulence. 
 In fact, these examples are simply the probabilistic classical field theory and statistical field theory analogs of quantum field theory.

\subsection{Semiclassical Evolution}
As far as the classical system is concerned, the dynamics of the complex phase factor (Eq.~\ref{eq:amp}) associated with the wavefunction is simply a choice of gauge.  In fact, there are alternate action principles that can be used to generate the Liouville equation \cite{Pfirsch91pfb, YeMorrison92pfb} and these formulations can also be used to define the dynamics of the phase factor of an associated KvN system.  

Perhaps the most physically relevant dynamics for the complex phase factor is determined by the semiclassical limit of the quantum wavefunction.  As is easily seen from Feynman's path integral formula, in this limit, the phase is given by the classical action.  The semiclassical phase factor,  originally derived by Van Vleck, must satisfy the Hamilton-Jacobi equation, and, a complete solution of the Hamilton-Jacobi equation  can be considered to be a function on phase space.  
 
Kostant's study of ``prequantization''  \cite{Kostant72preprint} was the first work to explicitly propose an equivalence between the classical action and the complex phase of the KvN wavefunction, as a function on phase space.  
Recently, Klein \cite{Klein18qsmf} suggested that the semiclassical phase might be of some importance and  Bondar, Gay-Balmaz, and Tronci, \cite{Bondar19prsa},  proposed that, in fact, it is the semiclassical dynamics that is physically relevant. These authors showed that the semiclassical KvN approach could be used to couple a semiclassical KvN system to a quantum system in a self-consistent manner.
%, for the first time.  
Note, however, that both Refs.~\cite{Klein18qsmf,Bondar19prsa} neglected to discuss the importance of the Maslov index \cite{Maslov72book, Maslov81book, Arnold67faia, Littlejohn92jsp} for obtaining the correct semiclassical branch of the phase factor.

The focus in this work is on quantum simulation of an arbitrary classical system of ODEs, and, developing a self-consistent framework for the semiclassical dynamics of Hamiltonian systems of ODEs will be left for future investigations.

\subsection{Quantum Simulation}
A quantum computer with finite resources can be used to simulate a finite-dimensional approximation of the KvN Hamiltonian operator. 
One must discretize the equations in a manner that can be represented with a finite number of qubits, which leads to a quantum mechanical coarse-graining and regularization of phase space that is rather different than that due to dissipative effects such as particle collisions.  In fact, one should squeeze the initial and final measurement states to achieve the uncertainty limit set by numerical discretization.

The KvN approach is analogous to the quantum simulation of the Schr\"odinger equation, where one obtains an approximate solution to the wavefunction at every time step and has the option of measuring the value of any given observable at the end of the simulation. Hence, it scales rather differently than the method of Ref. \cite{Leyton08arxiv} because the required memory does not grow exponentially with the number of time steps.  Moreover,  because one simulates the evolution of the PDF on phase space, one is effectively obtaining the solution for ``many trajectories'' at once.  

If the KvN Hamiltonian is sparse, e.g. local or banded, then the quantum representation of the classical system leads to exponential savings in the memory and computational work  \cite{Lloyd96sci} required for a \added{deterministic} Eulerian discretization of the Liouville equation.
\changed{Probabilisitic} time-dependent Monte Carlo (MC) algorithms \cite{KalosMCbook} can also provide a similar savings,
\added{
and it is important to compare the complexity of MC simulation to that of  KvN quantum simulation. 
}

\added{
The ``classical sampling'' strategy of averaging the outcome of multiple repeated measurements generally requires an amount of computational work that scales as one over the accuracy squared; i.e. $1/\varepsilon^2$, where $\varepsilon$ is the error measured in units of standard deviations.
However, the amplitude estimation algorithm \cite{Grover98arxiv, Brassard00arxiv}, which is the prototypical ``quantum sampling'' strategy,  achieves a quadratic speedup relative to  classical sampling by amplifying the amplitude of the desired state (an algorithm that can also be thought of as a type of quantum walk \cite{Szegedy04focs}). 
By utilizing amplitude estimation, a number of quantum algorithms \cite{Grover98arxiv, Abrams99arxiv, Brassard00arxiv, Heinrich01jc, Brassard11arxiv, Montanaro15prsa} can compute  numerical approximations to sums and integrals with a quadratic speedup relative to classical Monte Carlo algorithms, i.e. the accuracy generally decreases as the number of samples itself so that the computational work scales as $1/\varepsilon$, up to polylogarithmic factors. 
This savings can be substantial because obtaining a three digit improvement in accuracy would require $10^6$ repetitions of the classical algorithm vs. $10^3$ repetitions of the quantum algorithm.
An overview of the complexity of quantum and classical probabilistic algorithms for computing sums and integrals can be found in \cite{HeinrichNovak01arxiv}. 
In the typical cases of interest, a polynomial quantum speedup can be achieved that approaches a quadratic speedup in the limit of large dimensions or non-smooth integrands.
}

\added{
The KvN approach naturally allows one to utilize the algorithms specified by A. Montanaro in Ref. \cite{Montanaro15prsa} (which combine a number of seminal ideas from \cite{Abrams99arxiv, Brassard00arxiv, Heinrich01jc, Brassard11arxiv}) to calculate observables  with a quadratic speedup over classical Monte Carlo algorithms.
Efficient preparation of useful initial states is also an important objective. There are a number of algorithms based on quantum walks \cite{Aharonov02arxiv, Szegedy04focs, Wocjan09pra, Montanaro15prsa} that can be used to perform the preparation of useful states, such as Maxwell-Boltzmann and other equilibrium distributions, that also achieve a quadratic speedup over  classical  Markov Chain Monte Carlo (MCMC) algorithms.
Thus, the combination of these methods allows the quantum simulation of KvN classical dynamics to provide a quadratic speedup (up to polylogarithmic factors) over classical probabilistic algorithms.
}
 %Since quantum algorithms based on quantum walks can often provide a quadratic speedup over classical Monte Carlo algorithms \cite{Aharonov02arxiv, Montanaro15prsa}, it is of interest to determine whether similar gains would generally be expected for classical dynamics.
% Classical time-dependent Monte Carlo algorithms \cite{KalosMCbook} can also provide a similar savings.  Since quantum algorithms based on quantum walks can often provide a quadratic speedup over classical Monte Carlo algorithms \cite{Aharonov02arxiv, Montanaro15prsa}, it is of interest to determine whether similar gains would generally be expected for classical dynamics.

\subsection{Outline}  
The next section describes the derivation of the generalized Koopman-von Neumann representation of classical mechanics.  Section \ref{sec:HamiltonianDynamics} discusses the important special case of Hamiltonian dynamics, including canonical Hamiltonian systems, generalized Hamiltonian systems, and general variational systems.  Section \ref{sec:WaveActionPrinciple} relates the KvN Hamiltonian to an action principle for the wavefunction and uses Noether's theorem to derive a number of conservation laws for the system. Section \ref{sec:QuantumSimulation} discusses the steps necessary to obtain high-fidelity quantum simulation of a classical system.  Finally, estimates of the complexity of using the KvN approach to classical dynamics are discussed in Sec.~\ref{sec:Complexity}. A summary of the conclusions is presented in the final section.

{\it The Einstein summation convention is used throughout.}

%\emph{2. From Liouville to Schr\"odinger.---}
 \section{From Liouville to Schr\"odinger \label{sec:Derivation} }

 \subsection{Classical Dynamics on Phase Space}
 Consider the solution of a system of  $\xDim$  classical ODEs with coordinates, $\Vx=\setlist{\coord^1,\coord^2,\dots}\in \Reals^\xDim$, that evolve in time, $\time\in \Reals$, via
\begin{align}Ê\label{eq:ode}
\Vxdot:=d\Vx/d\time=\Vvel(\Vx,\time)
Ê\end{align}
where $\Vvel(\Vx,\time)$ is an arbitrary vector field.  

\begin{figure}[t]
\includegraphics[height=2.5in]{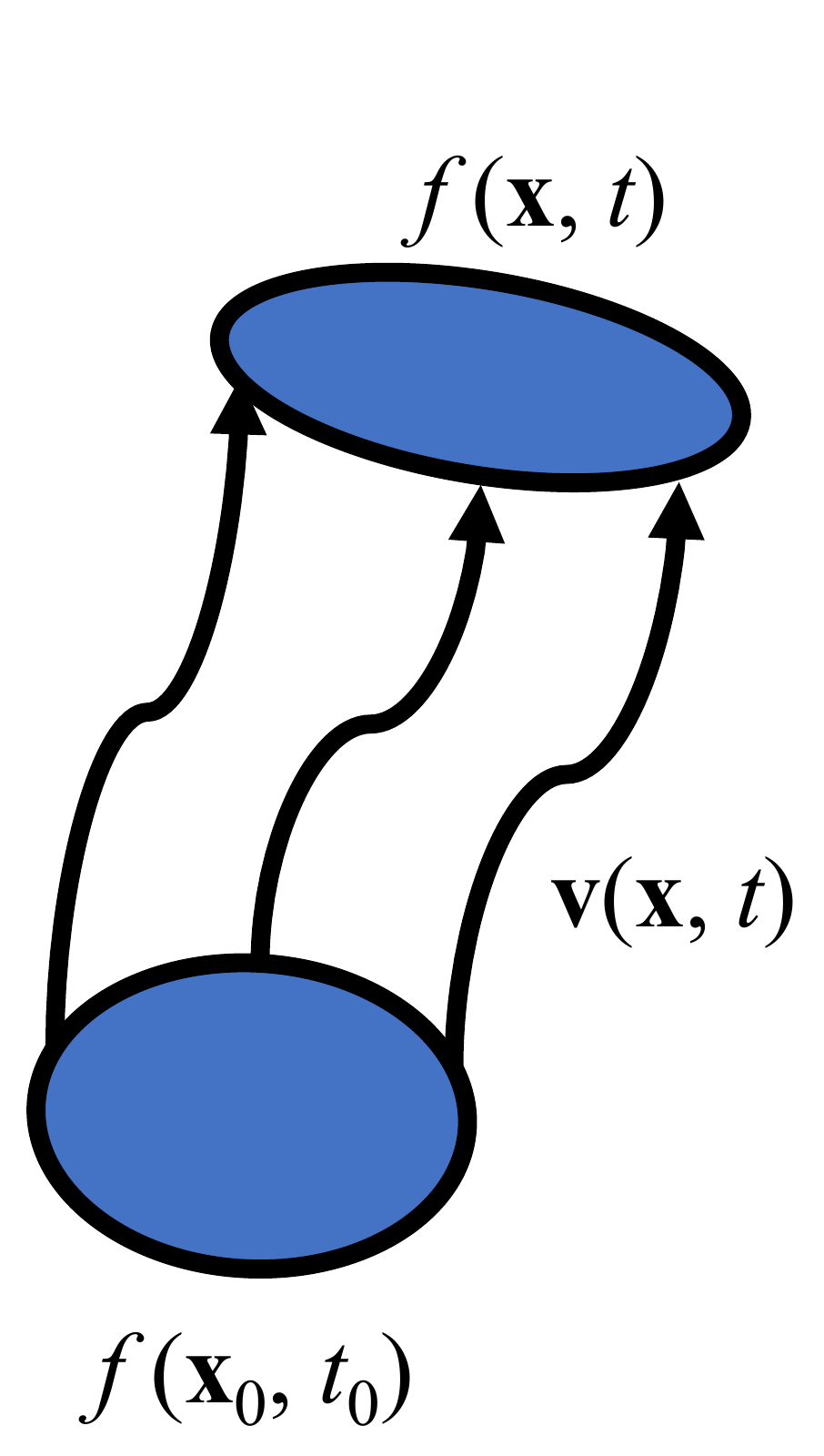}
\caption{The  probability distribution function (PDF), $\pdf(\Vx, \time)$, is advected by the flow $\Vvel(\Vx,\time)$, as described by the Liouville equation (Eq.~\ref{eq:liouville}).}
\label{fig:Liouville}
\end{figure}

There are two different ways to interpret the dynamics of observables or functions on phase space, as illustrated in Fig. \ref{fig:Liouville}. From the Lagrangian point of view, one can start with a number of initial conditions and follow the dynamics forward in time.  In this picture, observables evolve via the ``total'' time derivative $d/d\time$. From the Eulerian point of view, one can think of phase space as a manifold with coordinates given by $\Vx$. In this picture, observables evolve in time through partial differential equations (PDEs) on phase space. These two pictures are related through the equivalence between the total time derivative and the advection operator
\begin{align}\label{eq:advection_operator}
d/d\time = \partial_\time+\Vvel \cdot \Vnabla := \partial_\time+\vel^j \partial_j.
\end{align}

Any constant of the motion,  
$\phase(\Vx,\time)$, 
is simply advected along the flow by the advection operator
Ê\begin{align}Ê\label{eq:advection}
ÊÊÊÊÊÊÊÊÊÊ\dot \phase \eqdefÊ \partial_\time \phase + \Vvel\cdot  \Vnabla \phase=0
.
Ê\end{align}
The evolution operator, also known as the propagator or transfer operator, corresponding to this PDE is known as the Koopman operator \cite{CvitanovicChaosBook}.

The expectation value of an observable, $\Observable(\Vx,\time)$, is given by integration over the probability distribution function (PDF), $\pdf(\Vx,\time)$,
\begin{align}
\avg{\Observable} =\int \Observable \pdf d^\xDim\coord.
\end{align}
Thus, as explained in App.~\ref{sec:PDF_volume_form}, the PDF is a volume form or phase space density.
Conservation of phase space density
is described by the Liouville equation
Ê\begin{align}Ê\label{eq:liouville}
ÊÊÊÊÊÊÊÊÊÊÊÊ Ê\dot \pdf + \pdf  \Vnabla\cdot \Vvel= \partial_\time\pdf +\Vnabla\cdot     \left( \Vvel  \pdf \right)=0.
Ê\end{align}
The  propagator corresponding to this PDE is known as the Perron-Frobenius operator \cite{CvitanovicChaosBook}.  The advection of the PDF by the flow is illustrated schematically in Fig. \ref{fig:Liouville}. The meaning of the Liouville equation is that the PDF, $\pdf(\Vx,\time)$, is an invariant measure of the velocity field, $\Vvel(\Vx,\time)$, over space-time. 

Using integration by parts over  space-time
\begin{align}
\int \pdf (d/d\time)  \phase  d^\xDim \coord d\time = -\int \phase (d  /d\time)^\dagger \pdf  d^\xDim \coord d\time 
\end{align}
demonstrates that the  the Liouville operator that appears in (Eq.~\ref{eq:liouville}) is the anti-Hermitian adjoint of the  advection   operator 
\begin{align} \label{eq:liouville_operator}
-(d/dt)^\dagger =  \partial_\time+ \Vnabla\cdot \Vvel :=   \partial_\time+ \partial_j v^j.
\end{align}
Iff the velocity is divergence free, 
\begin{align}
0=\Vnabla\cdot \Vvel:=\partial_j v^j,
\end{align}
then the advection operator (Eq.~\ref{eq:advection_operator}) and the Liouville operator (Eq.~\ref{eq:liouville_operator})   are anti-Hermitian operators, i.e. $
\hat A~=~-~\hat A^\dagger$, and hence, identical. 
In the divergence-free case, the fact that these operators are anti-Hermitian implies that the corresponding evolution operator is unitary.

Discussion of the general form of the Liouville equation on space-time and the corresponding conventions for the divergence operator are discussed in Appendices \ref{sec:PDF_volume_form}-\ref{sec:PDF_space_time}.

\subsection{Koopman-von Neumann Hamiltonian \label{sec:KvN_Hamiltonian} }
The Koopman-von Neumann (KvN) approach to classical mechanics parallels the development of the postulates of quantum mechanics. The probability distribution function, $\pdf$, is defined as the inner product of the complex probability amplitude,  $\psi(\Vx,\time)$, with its adjoint, $\psi^\dagger(\Vx,\time)$,
\begin{align}  \label{eq:pdf=|amp|^2}
\pdf=\psi^\dagger\psi.
\end{align}
Thus, the probability amplitude or wave function
\begin{align}\label{eq:amp}
\amp=\pdf^{1/2}e^{i\phase},
Ê\end{align}
is closely related to notion of the ``square root'' of $\pdf$.  The expectation value of any phase-space observable,  $O(\Vx,\time)$, is given by integration over the probability distribution function (PDF), $\pdf(\Vx,\time)$,
\begin{align}\label{eq:General_KvN_Hilbert_Space_Inner_Product}
\left<O \right> =\int O \amp^\dagger\amp d^\xDim\coord = \int O  \pdf d^\xDim \coord.
\end{align}
This requires that the PDF is normalized to yield unit probability after integration over all of phase space.

Assume that the phase, $\phase$, satisfies the general equation of motion \cite{Klein18qsmf}
Ê\begin{align}Ê\label{eq:phase_dynamics}
ÊÊÊÊÊÊÊÊÊÊÊ \dot \phase = \partial_\time \phase + \Vvel\cdot  \Vnabla \phase= -\PhaseAction(\Vx,\time) /Ê\hbarÊ
.
Ê\end{align}
Then, taking the time derivative of Eq.~\ref{eq:amp} and using Eqs.~\ref{eq:liouville} and  \ref{eq:phase_dynamics} proves that $\psi$ satisfies the equation 
Ê\begin{align}Ê\label{eq:phase_dynamics}
ÊÊÊÊÊÊÊÊÊÊÊÊÊÊ \dot \amp +    \tfrac{1}{2}\amp\Vnabla\cdotÊ\Vvel
		 =\partial_\time \amp+ \tfrac{1}{2}\left(\Vvel \cdotÊ\Vnabla  +\Vnabla\cdotÊ\Vvel \right) \amp =-i\PhaseAction\amp /\hbar
		 .
Ê\end{align}
Simply multiplying this equation by $i\hbar$ yields the generalized Koopman-von Neumann (KvN) equation
\begin{align}Ê\label{eq:KvN_equation}
Ê i\hbar \partial_\time \amp &= -i\hbar\tfrac{1}{2}\left( \Vvel \cdotÊ\Vnabla  +\Vnabla\cdotÊ\Vvel   \right) Ê \amp+\PhaseAction\amp.
\end{align}
\added{The relationship between the KvN equation and the Schr\"odinger equation and the physical meaning of the $W$ function, which determines the phase factor, $\phase$, is discussed further in Sec. \ref{sec:semiclassical}.}

The KvN equation has the form of a Schr\"odinger equation 
\begin{align}
i\hbar \partial_\time \amp=Ê\Hhat \amp.
\end{align}
As can be proven using integration by parts, the KvN Hamiltonian operator, $\Hhat$, is Hermitian over the 
Hilbert space inner product defined on any two functions of phase space via
\begin{align}
\left<\phase | \amp\right> := \int \phase^\dagger(\Vx,\time) \amp(\Vx,\time) d^\xDim \coord.
\end{align} 
Hence, define the usual momentum operator, 
$\VPhat =-i\hbar \Vnabla$, 
and  position operator, 
$\Vxhat =\Vx$,  
  and promote any function of the coordinates to an operator, e.g. 
$\Vvhat=\Vvel(\Vxhat,\time)$ 
and 
$\hat \PhaseAction=\PhaseAction(\Vxhat,\time)$,
via the formal Taylor series expansion. Thus, one arrives at the generalized Koopman-von Neumann Hamiltonian operator
\begin{align}Ê\label{eq:KvN_Hamiltonian}
\Hhat &=  \tfrac{1}{2} \left(\VPhat\cdot \Vvhat  +   \Vvhat\cdot  \VPhat\right)  +\hat \PhaseAction .
\end{align}
Because the KvN Hamiltonian is Hermitian, the corresponding KvN evolution operator is unitary. 
 Koopman and von Neumann only considered the case of Hamiltonian dynamics, but the generalized form leads to a unitary evolution operator for any set of ODEs, even if they are not Hamiltonian.
 
In contrast to the usual Schr\"odinger equation, the KvN Hamiltonian is linear in momentum, $\VPhat$, because the KvN equation is a first order PDE rather than a second order PDE. Hence, there are a number of important differences between the behavior of the KvN equation and the usual Schr\"odinger equation as well as in the mathematical analysis of the two types of equations.  

The linear dependence on momentum implies that Heisenberg's equations of motion for the original phase space variables are exactly the same as the original classical equations of motion, 
\begin{align}
d\Vxhat/d\time=\Commutator{\Vxhat,\Hhat}/i\hbar= \Vvel(\Vxhat,\time).
\end{align} 
Thus, the $\Vxhat$ operators have the same solution as the original classical equations of motion.
In contrast, Heisenberg's equations of motion for the conjugate momenta  
\begin{align} \label{sec:Heisenberg_Pdot}
d\VPhat/d\time&=\Commutator{\VPhat,\Hhat}/i\hbar\\
&=-(\Vnabla \Vvhat)\cdot  \VPhat   +  \Vnabla \left( i\hbar \tfrac{1}{2} \Vnabla\cdot \Vvhat - \hat \PhaseAction \right)
\end{align} 
generically receive  a ``quantum correction'' when compared to the classical limit (Eq.~\ref{eq:momentum_eom}).

If there is a natural volume form on phase space, then the PDF and the wavefunction can also be  treated as scalar fields.  The Koopman-von Neumann approach for scalar fields  is discussed in App.~\ref{sec:PDF_scalar_form}. The Koopman-von Neumann equation on  space-time is discussed in App.~\ref{sec:PDF_space_time}.

\subsection{Koopman-von Neumann Evolution Operator \label{sec:KvN_Evolution}}
The phase space evolution described by Eqs.~\ref{eq:liouville}, \ref{eq:phase_dynamics}, and \ref{eq:KvN_equation} can be solved using the method of characteristics. The solution of the ODE's in Eq.~\ref{eq:ode}
has the form $\Vx=\Vxi(\Vx_0,\time)$ for the initial conditions $\Vx_0:=\Vx(\time_0)$. The inverse relation $\Vx_0=\Vxi^{-1}(\Vx,\time)$ allows one to express the solution of these equations in terms of the determinant 
\begin{align}
\Jac_0(\Vx_0,\time) =\left. \det \left(\partial \coord^j_0/\partial {\coord^k}  \right)\right|_{\Vx=\Vxi(\Vx_0,\time)}
.
\end{align}
The PDF is determined by
\begin{align}
\pdf(\Vx,\time) &= |\Jac_0(\Vx_0,\time) |\pdf(\Vx_0,\time_0)
\end{align}
and, if one defines
\begin{align}
\PhaseAction_0(\Vx_0,\time):=\PhaseAction({\mathbf \xi}(\Vx_0,\time),\time),
\end{align}
then the phase is determined by
\begin{align}
\phase(\Vx,\time)= \phase(\Vx_0,\time_0) -\int_{\time_0}^\time \PhaseAction_0(\Vx_0,\time) d\time/\hbar.
\end{align}
Thus, the evolution of the amplitude, $\amp$, can be determined from the definition in Eq.~\ref{eq:amp}
\begin{align} \label{eq:amp_vs_time}
\amp(\Vx,\time) &=   \Jac_0^{1/2}(\Vx_0,\time) e^{-i\int_{\time_0}^\time \PhaseAction_0(\Vx_0,\time) d\time/\hbar}  \amp(\Vx_0,\time_0)
.
\end{align}

The choice of the complex phase of $\Jac_0^{1/2}$ is irrelevant for the classical system; for example, one could formally use $\abs{\Jac_0}^{1/2}$ instead.  However, the phase shift due to the square root of the Jacobian, $\Delta \phase=-\nu \pi/2$, is given by the integer Maslov  index, $\nu$, which counts the number of zeros of the Jacobian along the trajectory since the starting point \cite{Maslov72book, Maslov81book, Arnold67faia, Littlejohn92jsp, CvitanovicChaosBook}. This phase shift must be correctly accounted for in order to obtain the correct semiclassical phase factor.  

The KvN evolution operator, $\EvolutionOp$, is defined to satisfy the operator equation
\begin{align} \label{eq:evolution_operator}
i\hbar \partial_\time \EvolutionOp=Ê\Hhat \EvolutionOp.
\end{align}
Using the general solution given above, the KvN evolution operator can be written as 
\begin{multline}
\left<\Vx \right| \EvolutionOp_{\time, \time_0} \left| \Vx_0\right>
\\
=\Jac_0^{1/2}(\Vx_0,\time)  \delta^\xDim\left(\Vx_0-\Vxi^{-1}(\Vx,\time)\right) e^{-i\int_{\time_0}^\time \PhaseAction_0(\Vx_0,\time) d\time/\hbar} 
\\
=\Jac_0^{-1/2}(\Vx_0,\time)  \delta^\xDim\left(\Vx-\Vxi(\Vx_0,\time)\right) e^{-i\int_{\time_0}^\time \PhaseAction_0(\Vx_0,\time) d\time/\hbar} 
.
\end{multline}
Again, because the generalized KvN Hamiltonian (Eq.~\ref{eq:KvN_Hamiltonian}) is Hermitian, the KvN evolution operator is unitary.

\subsection{Semiclassical Dynamics and the  Phase Factor  \label{sec:semiclassical} }
The dynamics of the phase factor, $\dot\phase=-\PhaseAction/\hbar$, has no effect on the classical dynamics. Within the confines of classical dynamics, the phase, $\phase$, is not measurable, and, hence, the choice of $\PhaseAction$ is equivalent to a choice of gauge. 

On the other hand, the semiclassical approximation to quantum dynamics represents the propagator as a sum over classical paths \cite{ Littlejohn92jsp,KleinertBook,CvitanovicChaosBook}, and, in this case, the phase factor is important for describing interference between paths. Similarly, when the classical system is coupled to a quantum system, then the dynamics is also  sensitive to the semiclassical phase factor introduced by the semiclassical system \cite{Bondar19prsa}.  In other words, Schr\"odinger's cat has rather different semiclassical phase factors depending on whether it is dead or alive, and, because the cat is entangled with the quantum part of the system, which may be in a superposition of states, this can lead to measurable interference effects due to the semiclassical phase of the cat itself. 

If the classical system is itself a Hamiltonian system (see Sec. \ref{sec:HamiltonianDynamics}), with the Hamiltonian, $\Ham(\qcoord^j,\pcoord_j,\time)$, specified as a function of generalized coordinates, $\qcoord^j$,  and conjugate momenta, $\pcoord_j$,  then there is a natural choice of phase factor  \cite{Kostant72preprint,Klein18qsmf,Bondar19prsa}.  If the phase velocity is determined by the classical Lagrangian, $\Lagrangian$,
\begin{align}\label{eq:semiclassical_phase}
\hbar\dot \phase = -\PhaseAction =\Lagrangian:=\pcoord_j \partial_{\pcoord_j}\Ham-\Ham,
\end{align}
then the phase factor is equal to the classical action and corresponds to the  semiclassical phase. The semiclassical phase factor is related to the momentum through the Eikonal approximation, 
\begin{align}
\left. \hbar \partial_{\qcoord^j} \phase \right|_{q_0,\time}=\pcoord_j  ,
\end{align}
and satisfies the Hamilton-Jacobi equation,
\begin{align}\label{eq:Hamilton-Jacobi}
\left.\hbar\partial_\time \phase\right|_{q_0,q}  = -\Ham(\qcoord^j,\pcoord_j, \time).
\end{align}
Here, the partial derivatives are taken as constant with respect to the initial value of the coordinates, $q^j_0$, or some other constants of the motion. Taking the total time-derivative of $\phase$ yields Eq. \ref{eq:semiclassical_phase}.

For short time intervals, only a single classical path will generically contribute to the propagator. However, for longer time intervals, multiple classical paths can contribute, each with a separate phase.
In order to obtain the semiclassical phase factor for each path, one must take care to ensure that the Maslov index \cite{Maslov72book, Maslov81book, Arnold67faia} is correctly accounted for, by choosing the correct phase of the square root of the Jacobian, $\Jac_0^{1/2}$, in Eq.~\ref{eq:amp_vs_time}. 
This generates an additional $- \pi/2$ phase jump whenever $\Jac_0$ passes through zero and leads to an additional overall phase shift, $\Delta \phase=-\nu \pi/2$, 
where $\nu$ counts the number of times that $\Jac_0$ passes through zero along the trajectory \cite{Littlejohn92jsp, CvitanovicChaosBook}. In other words, as the Jacobian passes through zero, the  square root periodically follows the sequence of branches: $(1,-i,-1,i)$.
(This important fact appears to have been neglected in the discussion of Refs.  \cite{Klein18qsmf,Bondar19prsa}.)

It is also possible to give the amplitude additional index structure, $\amp_{ijk\dots}$, so that the various components transform as a representation of the Poincar\'e group under changes of reference frame. Clearly, this extension is important for describing the semiclassical evolution of higher spin fields as well as for describing interactions with gauge fields. While this does not affect the classical dynamics, the semiclassical dynamics could certainly be an interesting avenue to pursue in   future work.

In the more general non-Hamiltonian case considered here, a natural choice of $\dot\phase$  is not immediately obvious. The trivial choice, $\PhaseAction=0$,   corresponds to the constrained  classical action (Eq.~\ref{eq:action_constrained}) considered in the next subsection.

The general case is also related to the quantum mechanics of a charged particle in an $\Ncoord$-dimensional vector potential, $\VAfield(\Vx,\time)$, and scalar potential, $\Phi(\Vx,\time)$,  in the limit that the canonical momentum is much smaller than the vector potential, $\norm{P} \ll \norm{\charge A}$, where $\charge$ is the charge of the particle. In this limit,  one can identify the relations, 
\begin{align}
\mass\Vvel &=-\charge\VAfield & \PhaseAction=\mass\vel^2/2+\charge \Phi,
\end{align}
where $\mass$ is the mass of particle.   Thus, in this limit, the choice of $\hbar \dot\phase=-\PhaseAction $ is equivalent to a choice of scalar potential, $\Phi$.

%\emph{3. The Constrained Hamiltonian.--}
\subsection{Constrained Hamiltonian \label{sec:Constrained_Hamiltonian}}
\begin{figure}[t]
\includegraphics[width=2in]{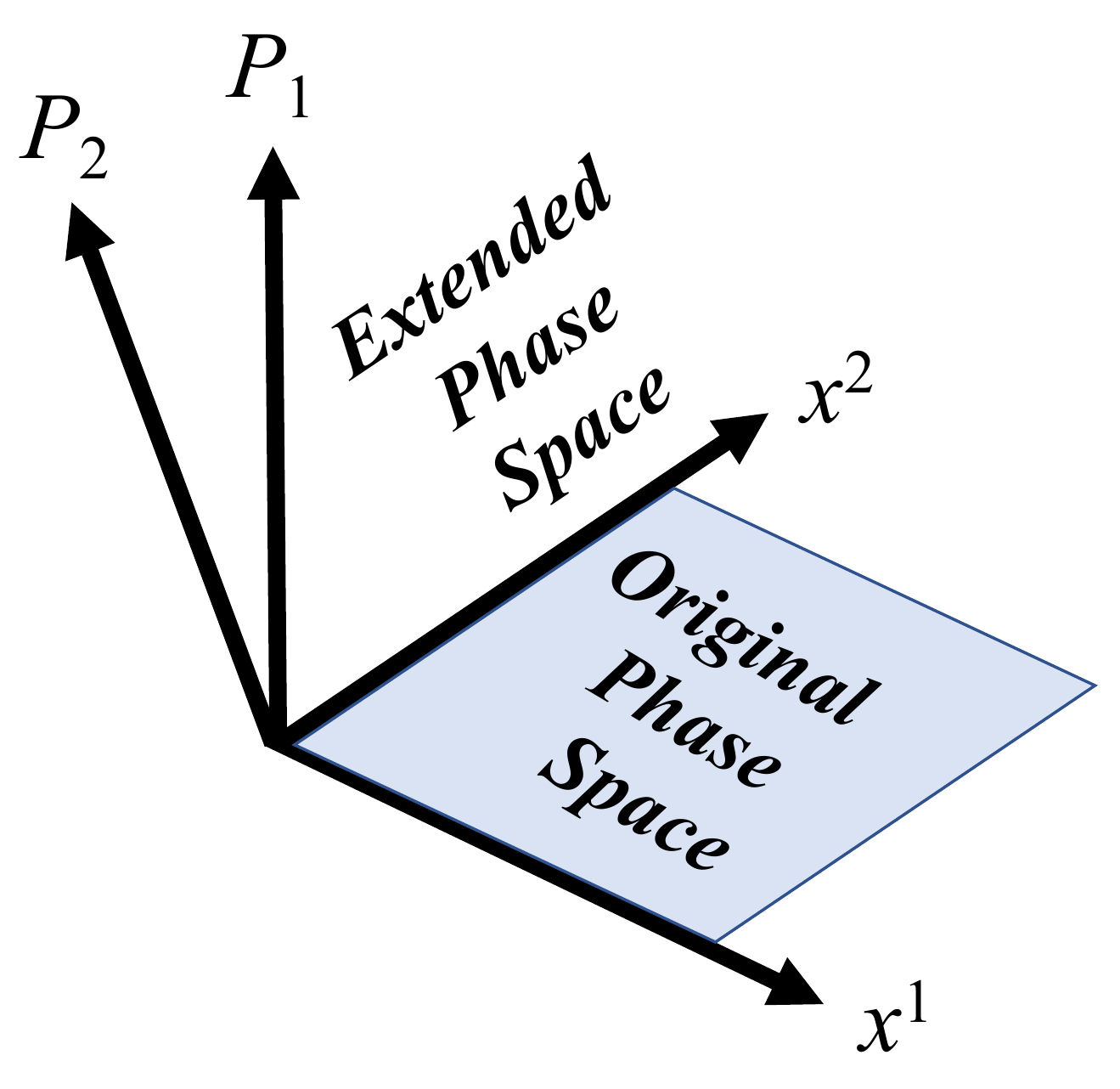}
\caption{The constrained Hamiltonian, $\CH$, (Eq.~\ref{eq:hamiltonian_constrained}) is defined on an extended phase space that includes the original phase space coordinates, $\Vx=\setlist{\coord^1,\coord^2,\dots } $, as well as the  canonically conjugate momenta, $\VP=\setlist{\Mom_1,\Mom_2,\dots }$, which act as Lagrange multipliers. }
\label{fig:extended_phase_space}
\end{figure}

In the limit $\hbar\rightarrow 0$,  the KvN Hamiltonian of Eq.~\ref{eq:KvN_Hamiltonian} becomes the classical Hamiltonian
Ê\begin{align}Ê\label{eq:hamiltonian_constrained} 
	   \Hconstraint(\Vx,\VP,\time) &:=\VP\cdot  \Vvel(\Vx,\time)+\PhaseAction(\Vx,\time).
\end{align}
Because it is linear in the momentum, $\VP$, this corresponds to a constrained Hamiltonian system.
Any set of $N$ classical ODE's can  be generated by using the action principle corresponding to this Hamiltonian \cite{Pontryagin62book}   
Ê\begin{align}Ê\label{eq:action_constrained} 
ÊÊÊÊÊÊÊÊÊÊÊÊ\CS[\Vx,\VP;\time] &:=\int \left[\VP\cdot  \left(\Vxdot  - \Vvel \right)-\PhaseAction\right] d\time  
ÊÊ\end{align}
which can be interpreted as as a sum over constraints.
Variation with respect to the Lagrange multipliers, $\Mom_j$, enforces the classical equation of motion for the  coordinates, $\coord^j$. Variation with respect to the coordinates, generates the classical equation of motion for the Lagrange multipliers 
\begin{align}Ê\label{eq:momentum_eom}
\dot \VP Ê= -( \Vnabla \Vvel )\cdot \VP  -\Vnabla\PhaseAction
\end{align}
(compare to Eq. \ref{sec:Heisenberg_Pdot}).
The equations of motion  ensure that the coordinate transformation   between different points in time always remains symplectic, and, in fact, canonical.

The constrained Hamiltonian can be quantized via any acceptable quantization procedure that reduces to Eq.~\ref{eq:hamiltonian_constrained}   in the limit $\hbar\rightarrow0$. The symmetric Weyl quantization rule leads to $\Hhat$ in Eq.~\ref{eq:KvN_Hamiltonian}. 
The constrained Hamiltonian in Eq.  \ref{eq:hamiltonian_constrained} implies that the quantized variables $\hat \coord^i$ commute with one another.  In contrast, other types of constraints, e.g. holonomic constraints, would require modifying the Poisson bracket to the Dirac bracket \cite{Dirac50cjm}. This would lead to nontrivial commutators and nontrivial uncertainty relations which could potentially cause deviations from the exact classical dynamics.

%\emph{5. Examples.---}
\subsection{General Examples \label{sec:General_Examples}}
Consider a single autonomous ODE in a single variable, $\coord$,
Ê\begin{align}Ê\label{eq:1d_ode}
	\dot \coord = \Vel(\coord).
Ê\end{align}
The KvN Hamiltonian operator is simply
Ê\begin{align}Ê
	\Hhat =  \tfrac{1}{2} \left(\Phat \VelOp +  \VelOp \Phat \right)=-i\hbar\left(\vel \partial_x + \tfrac{1}{2} \vel'\right) .
Ê\end{align}
The equation of motion can be derived from the constrained Hamiltonian, $\Hconstraint=\Pconstraint\Vel $. In addition to Eq.~\ref{eq:1d_ode}, the canonical momentum satisfies the equation
Ê\begin{align}Ê
	\dot  \Pconstraint &= -  \Pconstraint \Vel'(\coord)  .
Ê\end{align}
The classical solution to this equation is simply
Ê\begin{align}Ê
	 \Pconstraint = \Pconstraint_0 \Vel_0/\Vel(\coord).
Ê\end{align}
Since these equations are symplectic, the phase space area is conserved, as can be checked directly from the equations of motion. Heisenberg's equation of motion is
Ê\begin{align}Ê
	d  \Phat /d\time &= - \tfrac{1}{2} \left(\Phat \VelOp'  +\VelOp'\Phat \right) =-\VelOp'\Phat+i\hbar\VelOp''.
Ê\end{align}

A specific instance of this case is  given by $\Vel=\gamma \coord$, where the sign of $\gamma$ determines whether the motion exhbits exponential growth or damping. The constrained Hamiltonian is simply $\Hconstraint = \Pconstraint  \gamma \coord$ and 
the classical solutions on extended phase space are 
Ê\begin{align}Ê
	 \coord&= \coord_0 e^{\gamma \time} &
	\Pconstraint& = \Pconstraint_0 e^{-\gamma \time}.
Ê\end{align}
The  corresponding KvN Hamiltonian is 
Ê\begin{align}Ê
	\Hhat  =  \tfrac{1}{2}\gamma \left(\Phat   \coordOp +  \coordOp \Phat\right) = -i\hbar \gamma \left(\coord \partial_\coord  +\tfrac{1}{2}  \right) .
Ê\end{align}

Consider the general linear equation set 
\begin{align}Ê
 \Vxdot=\Amat \cdot \Vx
 .
Ê\end{align}
The constrained Hamiltonian 
\begin{align}Ê
	\Hconstraint 	=\VP^\dagger\cdot \Amat\cdot \Vx
Ê\end{align}
leads to the additional equations of motion,  
\begin{align}Ê
\dot \VP  =-\Amat^\dagger \cdot \VP
.
Ê\end{align}
The corresponding KvN Hamiltonian is given by
Ê\begin{align}Ê\label{eq:quantum_ham}
	\Hhat   &=
 \tfrac{1}{2}\left( \VPhat^\dagger\cdot \Amat\cdot \Vxhat + \Vxhat^\dagger\cdot \Amat\cdot \VPhat\right) 
 \\
 &=-i\hbar\left(\Vxhat^\dagger\cdot \Amat\cdot \nabla+\tfrac{1}{2} \tr{\Amat}\right)
 .
Ê\end{align}

Since the last two examples had linear equations of motion resulting from Hamiltonians that are quadratic forms, Heisenberg's equations of motion are the same as the classical equations.
The Hermitian block off-diagonal quadratic form of the Hamiltonian leads to a symplectic block diagonal form for the equations of motion.

%\emph{4. Hamiltonian Dynamics.--}
\section{  Hamiltonian Dynamics \label{sec:HamiltonianDynamics}}
\subsection{Canonical Hamiltonian Systems \label{sec:Canonical_Hamiltonian} }
 Hamiltonian systems of differential equations are of fundamental  importance to physics and mathematics. 
Canonical coordinates are defined by pairs of configuration space coordinates, $q^j$, and the corresponding conjugate momenta, $p_j$.
In canonical coordinates, Hamilton's equations take the canonical form
 Ê\begin{align}Ê
	\dot q^j&=   \partial_{p_j}    H &  \dot p_j&= -  \partial_{q^j}    \Ham 
	.
Ê\end{align}

Hamilton's equations are divergence-free in canonical coordinates, which implies that the natural measure is the canonical measure $d^\xDim q d^\xDim p$.
The expectation value of any phase-space observable,  $O(q^j,p_k,\time)$, is given by integration over the probability distribution function (PDF), $\pdf(q^j,p_k,\time)=\psi^\dagger\psi$,
\begin{align}\label{eq:Hamiltonian_KvN_Hilbert_Space_Inner_Product}
\left<O \right> =  \int O  \psi^\dagger\psi d^\xDim\qcoord d^\xDim\pcoord.
\end{align}
Similarly, the inner product of any two functions is defined via
\begin{align}\label{eq:Hamiltonian_KvN_Hilbert_Space_Inner_Product}
\left<\phase|\amp \right> =\int \phase^\dagger\amp    d^\xDim\qcoord d^\xDim\pcoord.
\end{align}

Since Hamilton's equations are divergence-free, the Liouville equation 
\begin{align}Ê
\dot \pdf = \partial_\time \pdf + \PoissonBracket{\pdf}{\Ham}=0
\end{align}
is equivalent to its space-time adjoint, advection by the flow. Here, the canonical Poisson bracket is defined as 
\begin{align}Ê
 \PoissonBracket{a}{b}=\partial_{q^j}a \partial_{p_j} b - \partial_{p^j}a \partial_{q_j} b.
\end{align}

Let us now introduce a new convention for the Lagrange multipliers $\setlist{P_j,Q^k}$ so that they are conjugate to the $\setlist{q^j,p_k}$ in the ``canonical'' manner
\begin{align}Ê\label{eq:KvN_PoissonBracket}
	\PoissonBracket{q^j}{P_k}&=\delta^j_k  & \PoissonBracket{Q^j}{p_k}&=\delta^j_k. 
Ê\end{align}
The constrained Hamiltonian takes the form
 \begin{align}Ê
	\Hconstraint &=(P_j  \partial_{p_j}  + Q^j   \partial_{q^j})   \Ham.
Ê\end{align}
The equations of motion of the Lagrange multipliers are 
 Ê\begin{align}Ê
	\dot Q^j&=    \left( P_k  \partial_{p_k}  + Q^k   \partial_{q^k} \right)\partial_{p_j}     \Ham  \\
	\dot P_j&=  - \left( P_k  \partial_{p_k}  + Q^k   \partial_{q^k} \right)  \partial_{q^j}     \Ham  .
Ê\end{align}

The usual Dirac quantization procedure for Hamiltonian systems promotes the classical Poisson bracket to commutation relations. Instead, the KvN approach only promotes the Poisson brackets of the Lagrange multipliers (Eq. Ê\ref{eq:KvN_PoissonBracket}) to commutation relations via the  definitions
\begin{align}
\Phat_j&=-i\hbar \partial_{\qcoord^j} & \Qhat_j=i\hbar \partial_{ \pcoord^j}.
\end{align}
Since the canonical equations of motion are divergence-free, the   KvN Hamiltonian  is simply
 Ê\begin{align}
	\Hhat  =   \Phat_j( \partial_{p_j}\hat \Ham )   +  \Qhat^j(\partial_{q^j}\hat\Ham )  +\hat\PhaseAction.
Ê\end{align}
In this formulation, the Heisenberg uncertainty principle only applies to the pairs $\setlist{q^j,P_j}$ and $\setlist{p_j,Q^j}$, but not to the pairs of $\setlist{q^j,p_j}$.

%\emph{5. Hamiltonian Examples.---}
 \subsection{Canonical Hamiltonian Examples \label{sec:Hamiltonian_Examples} }
Consider the classical harmonic oscillator, with HamiltonianÊ
\begin{align}Ê
	\Ham =  \omega_0 (q^2+p^2)/2 .
Ê\end{align}
The dynamics can be exactly represented by the constrained Hamiltonian
Ê\begin{align}Ê
	\Hconstraint = \omega_0 (p P  + q Q )  . 
Ê\end{align}
The additional nontrivial equations of motion for the Lagrange multipliers are
Ê\begin{align}Ê
	\dot Q  &= \omega_0 P  & 	\dot P  &= -\omega_0  Q .
Ê\end{align}
Hence, the dynamics on the extended phase space consists of two harmonic oscillators which rotate in the same direction.  
The quantum Hamiltonian is  
Ê\begin{align}Ê
	\Hhat  =   \omega_0(p \Phat + q\Qhat )=-i\hbar\omega_0 \partial_\angle .
Ê\end{align}
where $\tan(\angle) = -p/q$.

Consider the dynamics of an integrable system of nonlinear oscillators defined by the Hamiltonian $\Ham_0(\action_j)$, where $\action_j$ are the conserved action variables corresponding to each oscillator. The evolution of the conjugate phases, $\angle^j$, is determined by the frequencies  $\omega_0^j := \partial_{\action_j} \Ham_0$. 
If one defines the number operators as
\begin{align}
\Nhat_j=- i\partial_{\angle^j}
\end{align}
then, this leads to   the  KvN Hamiltonian 
Ê\begin{align}Ê
	\Hhat  =\hbar \omega_0^j \Nhat_{ j}   .
Ê\end{align}
While the number operators are conserved like the action coordinates, the angle operators defined by 
\begin{align}
\Anglehat^j=  i\hbar\partial_{\action_j}
\end{align}
satisfy the Heisenberg equations  of motion
\begin{align}
d\Anglehat^j/d\time=  \Nhat_{ k} \partial_{\action_j}\hbar\omega_0^k.
\end{align}

\subsection{Generalized Hamiltonian Systems \label{sec:General_Hamiltonian}}
In  general, a  Hamiltonian system  of equations is defined both by the  Hamiltonian $\Ham(\coord,\time)$ and by the Poisson bracket \cite{Olver93book,Morrison98rmp}
\begin{align}
\PoissonBracket{a}{b}=  \PoissonTensor^{jk} \partial_j a \partial_k b.
\end{align} 
The Poisson 2-vector, $ \PoissonTensor^{jk}(\coord,\time)$,  must be an antisymmetric tensor and must satisfy the Jacobi identity
\begin{align}
0=\PoissonBracket{\PoissonBracket{a}{b}}{c} + \PoissonBracket{\PoissonBracket{b}{c}}{a}+ \PoissonBracket{\PoissonBracket{c}{a}}{b}
\end{align} 
which yields the relation
\begin{align}
\PoissonTensor^{il} \partial_l \PoissonTensor^{jk} +\PoissonTensor^{jl} \partial_l \PoissonTensor^{ki}+\PoissonTensor^{kl} \partial_l \PoissonTensor^{ij}=0.
\end{align}
The classical equations of motion generated by the Hamiltonian $\Ham(\coord,\time)$ are
Ê\begin{align}Ê\label{eq:EOM_generalH}
	\dot \coord^j =\vel^j=\PoissonBracket{\coord^j}{\Ham}=\PoissonTensor^{jk} \partial_k \Ham .
Ê\end{align}

Following the general procedure outlined previously, the corresponding  constrained Hamiltonian is
Ê\begin{align}Ê\label{eq:hamiltonian_Hconstrained}
	\Hconstraint &=    \Pconstraint_j \PoissonBracket{\coord^j}{ \Ham } = \Pconstraint_j \PoissonTensor^{jk} \partial_k \Ham+\PhaseAction
	.
Ê\end{align}
This leads to the following dynamics of the Lagrange multipliers:
 Ê\begin{align}Ê
	\dot \Pconstraint_i&=  -\Pconstraint_j  \partial_i  \left(\PoissonTensor^{jk} \partial_k \Ham\right)  -\partial_j\PhaseAction
	.
Ê\end{align}
The  KvN Hamiltonian for generalized Hamiltonian systems is
Ê\begin{align}Ê
	\Hhat &=  \tfrac{1}{2} \left( \Phat_j \velhat^j + \velhat^j\Phat_j \right) + \hat \PhaseAction
	\\
	&=\tfrac{1}{2}\left( \Phat_j \PoissonTensor^{jk} \partial_k\HamOp + \PoissonTensor^{jk} \partial_k\HamOp \Phat_j  \right)+\hat \PhaseAction
	\\
	&=	 \tfrac{1}{2}\left( \Phat_j \PoissonBracket{\coordOp^j}{\HamOp} +  \PoissonBracket{\coordOp^j}{\HamOp}  \Phat_j \right)+\hat \PhaseAction
		  \label{eq:hamiltonian_HKvN}
.
\end{align}
Once again, this form follows the convention for general systems of differential equations, but does not correspond to the convention used for  canonical Hamiltonian systems used in Sec.~\ref{sec:Canonical_Hamiltonian}.

When the generalized Hamilton's equations of motion are non-singular, they have a canonical definition of the volume form.  Thus, as explained in App.~\ref{sec:PDF_canonical_volume_form}, one can also derive a corresponding KvN equation where the PDF and the wavefunction are treated as scalar fields.

\subsection{General Variational Systems \label{sec:Variational_Hamiltonian}}
A  general variational system of equations is defined by extremizing the action principle \cite{Olver93book}
\begin{align}
\CS[\coord;\time]=\int \left[\alpha_j(\coord,\time)  d\coord^j-\Ham(\coord,\time ) d\time\right].
\end{align}
The action is defined in terms of the Poincar\'e 1-form, $\PoincareForm=\alpha_j \Vd \coord^j$. (For an introduction differential forms and exterior calculus, see Refs.~\cite{SchutzBook, AbrahamMarsdenBook}.)
This leads to the equations of motion 
\begin{align} \label{eq:variational_EOM}
\LagrangeTensor_{jk}  \dot \coord^k = \partial_j\Ham + \partial_\time\alpha_j,
\end{align}
where the symplectic 2-form, $\LagrangeForm:=\Vd\PoincareForm$, which defines the Lagrange bracket, is the exterior derivative of the Poincar\'e 1-form; i.e.  in components 
\begin{align}
\LagrangeTensor_{jk} = \partial_j \alpha_k-\partial_k \alpha_j.
\end{align}
 These equations can be written in the form of Hamilton's equations of motion 
\begin{align}
 \dot\coord^j= \vel^j:=\PoissonTensor^{jk}    \left(\partial_k \Ham + \partial_\time\alpha_k \right).
\end{align}
with the identification of the Poisson tensor as the inverse of the Lagrange form, $\PoissonForm=\LagrangeForm^{-1}$.  The  Jacobi identity follows from the fact that the symplectic 2-form is closed, since it is exact $\Vd\LagrangeForm=\Vd^2\PoincareForm=0$. 

\def\Vpartial{{\boldsymbol \partial}}
The canonical equations of motion result from the Poincare 1-form, $\PoincareForm=p_j\Vd q^j$. This yields the canonical symplectic 2-form
\begin{align}
\LagrangeForm=\Vd p_j\wedge \Vd q^j
\end{align}
and the canonical Poisson 2-vector 
\begin{align}
\PoissonForm = \Vpartial_{q^j} \wedge \Vpartial_{p_j}  .
\end{align}
Generalized Hamiltonian systems and general variational systems are equivalent when the symplectic 2-form, $\LagrangeForm$, is non-degenerate.  For example, Hamilton's equations in Eq.~\ref{eq:EOM_generalH} result when $\partial_\time\PoincareForm=0$. Alternatively, the general variational equations in Eq.~\ref{eq:variational_EOM} can be embedded within Hamilton's equations on extended phase space, with the additional coordinates $\setlist{\tau,p_\tau}$ and the Hamiltonian $\bar \Ham = \Ham+p_\tau$.

The variational equations of motion (Eq.~\ref{eq:variational_EOM}) result from the constrained Hamiltonian
\begin{align} \label{eq:variational_Hconstrained}
\Hconstraint= \Coord^j  \left(\partial_j \Ham + \partial_\time\alpha_j \right) + \PhaseAction =  \Coord^j  \LagrangeTensor_{jk}\vel^k+\PhaseAction
\end{align}
where, now, the  $\Coord^j$ are the Lagrange multipliers that enforce the variational equations of motion.
The action principle is now
\begin{align} \label{eq:variational_action}
\CS[\coord,\Coord;\time]=\int \left[ \Coord^j  \LagrangeTensor_{jk} \left(\dot \coord^k-\vel^k \right) - \PhaseAction\right]d\time.
\end{align}
The Hamiltonian form in Eq.~\ref{eq:hamiltonian_Hconstrained} and the variational form in Eq.~\ref{eq:variational_Hconstrained} are equivalent with the definition
\begin{align}
\Coord^k= \Pcoord_j \PoissonTensor^{jk} .
\end{align}
Due the fact that the Poisson tensor is conserved by the equations of motion, $d\LagrangeForm/dt=0$  (proven in App. \ref{sec:symplectic_form_conservation}), the Lagrange multipliers, $\Coord^j$, satisfy the equations of motion
\begin{align} \label{eq:EOM_Hconstrained}
\dot \Coord^j= -\PoissonTensor^{jk} \left[\Coord^l  \partial_k \left(\partial_l \Ham + \partial_\time\alpha_l \right)+\partial_k \PhaseAction\right].
\end{align}
The analogous quantized operators, $\Coordhat^j$, should be given a Hermitian form
\begin{align}
\Coordhat^k=\tfrac{1}{2} \left( \hat \PoissonTensor^{jk} \Phat_j+ \Phat_j  \hat \PoissonTensor^{jk} \right).
\end{align}
Thus, the  KvN Hamiltonian for general variational systems is 
\begin{align}
\Hhat&= \tfrac{1}{2} \left( \Phat_j \velhat^j + \velhat^j\Phat_j \right) + \hat \PhaseAction
\\&= \tfrac{1}{2} \left( \Coordhat^j  \hat \LagrangeTensor_{jk}\velhat^k + \velhat^k \hat \LagrangeTensor_{jk} \Coordhat^j \right) + \hat \PhaseAction.
 \label{eq:variational_HKvN}
\end{align}
The final form corresponds to the convention that was used for canonical Hamiltonian systems in Sec.~\ref{sec:Canonical_Hamiltonian}.

For general variational systems, there is a canonical volume form induced by the symplectic 2-form. Thus, as explained in App.~\ref{sec:PDF_canonical_volume_form}, one can also derive a corresponding KvN equation where the PDF and the wavefunction are treated as scalar fields.

%\emph{6. Wave Action Principle.---}
\section{Wave Action Principle and Conservation Laws \label{sec:WaveActionPrinciple}} 
\subsection{Action Principle for General Systems}
The   Lagrangian that generates the KvN equation for the wave function, Eq.~\ref{eq:phase_dynamics}, as well as the Liouville equation and its adjoint, Eqs.~\ref{eq:advection} and \ref{eq:liouville}, is given by
\begin{align}
{\mathfrak L} 
&  =  \int  \amp^\dagger \left(i\hbar\partial_\time-\Hhat\right) \amp   d^\xDim \coord
\\
  &= -\int \pdf \left(\hbar\dot \phase +\PhaseAction\right)d^\xDim \coord
\end{align}
where integration by parts is used to derive the final expression.
The Hamiltonian that generates these partial differential equations is then given by
\begin{align}
{\mathfrak H}
 &=  \int  \amp^\dagger \Hhat  \amp d^\xDim \coord \\
 &=   \int \left[-i  \tfrac{\hbar}{2}\amp^\dagger \Vvel\cdot\Vnabla  \amp -i\tfrac{\hbar}{2} \amp^\dagger  \Vnabla\cdot\Vvel  \amp+\PhaseAction  \amp^\dagger\amp \right] d^\xDim \coord \\
 &=  \int \left[ \hbar\pdf \Vvel\cdot\Vnabla   \phase+\PhaseAction  \pdf  \right] d^\xDim \coord 
 .
 \label{eq:Wave_Hamiltonian_grad}
\end{align}
Again, integration by parts is used to derive  the final expression.
The definition of the Hamiltonian also allows one to derive the expression
\begin{align}\label{eq:Wave_Hamiltonian_ddt}
{\mathfrak H}
 &= \Im \int \amp^\dagger i\hbar\partial_\time \amp d^\xDim\coord =-\int \pdf\hbar \partial_\time \phase d^\xDim \coord.
\end{align}

There are a number of additional action principles that have been discussed in the literature \cite{Pfirsch91pfb, YeMorrison92pfb, Bondar19prsa}. After conversion to Koopman-von Neumann form, these action principles can be interpreted as providing an alternate dynamics of the phase factor. Hence, it may be interesting to explore the KvN dynamics that is implied using these alternate formulations as well.  

\subsection{Action Principle for Canonical Hamiltonian Systems}
For the case of a canonical Hamiltonian system, with Hamiltonian, $\Ham(\qcoord^j,\pcoord_j,\time)$, integration by parts  allows one to express the Hamiltonian in the  form  
\begin{align}
{\mathfrak H}  &=\int \left[ -i\hbar   \PoissonBracket{ \amp^\dagger}{\amp }\Ham  + \PhaseAction \amp^\dagger\amp \right]d^\xDim\coord
\\
&= \int \left[  \hbar \PoissonBracket{ \pdf}{\phase} \Ham+\PhaseAction \pdf \right] d^\xDim \coord.
\end{align}

With the gauge choice $\hbar\dot\phase=-\PhaseAction=0$, the probability density  that determines the KvN Hamiltonian density, $\Ham \pdf_\Ham$, is
\begin{align}
\pdf_\Ham=- i\hbar \PoissonBracket{ \amp^\dagger}{\amp} =  \PoissonBracket{\pdf}{\hbar \phase}.
\end{align}
This will only be equal to $\pdf$, iff the phase $\phase$ is canonically conjugate to  $\ln{(\pdf)}$
\begin{align} \label{eq:PB_pdf_phi}
  \PoissonBracket{\ln{(\pdf)}}{\hbar \phase} =1.
\end{align} 
Due to the Jacobi identity, this condition is preserved by the dynamics, so that if Eq.~\ref{eq:PB_pdf_phi} is true as an initial condition, then it is true for all time. 
The same is true if both $\pdf$ and $\PhaseAction$ are functions of adiabatic invariants alone, in which case the Hamiltonian density becomes $ (\Ham+\PhaseAction)\pdf$.

For example, consider the $\xDim$-dimensional Maxwellian distribution function
\begin{align}
\pdf = \exp{(- \pcoord^2/2 \mass T)}/(2\pi \mass T)^{\xDim/2}
\end{align}
where $\pcoord_i=\mass g_{ij} \vel^j$ and $T$ is the temperature. A family of solutions for the conjugate phase factor is given by
\begin{align}
\hbar \phase =  ( k_j \qcoord^j/ k_i \vel^i)T .
\end{align}
More generally, consider action-angle coordinates $\setlist{\action_j,\angle^j}$, where $\dot \angle^j=\omega^j(\action)=\partial_{\action_j}\Ham$.
If the PDF is a function of the action alone, then a solution for the phase is 
\begin{align}
\hbar \phase =  k_j \angle^j/   k_i \partial_{\action_i} \log{(\pdf)},
\end{align}
and, if the PDF is a function of the Hamiltonian alone, $\pdf(\Ham)$, then
\begin{align}
\hbar \phase =  (k_j \angle^j/ k_i \omega^i) d\Ham/d \log{(\pdf) }.
\end{align}

In the canonical case, the Poisson bracket can also be expressed as a divergence, so that
\begin{align} \label{eq:PB_pdf_divergence}
\pdf_\Ham= \partial_j \left(\pdf J^{jk}\hbar \partial_k\phase\right),
\end{align}
and this form of $\pdf_\Ham$ as a divergence is also correct for generalized Hamiltonian dynamics.
 Ref.~\cite{Bondar19prsa}  claimed that Eq.~\ref{eq:PB_pdf_phi} is not compatible with the fact that $\pdf_\Ham$ can be written as a divergence, Eq.~\ref{eq:PB_pdf_divergence},
 and, thus, as a surface integral, which they assumed can be made to vanish at the boundary. However, in order to satisfy  Eq.~\ref{eq:PB_pdf_phi}, the phase function, $\phase$, must be multiple valued, so that the surface integral does not vanish.

Alternatively, assume that the semiclassical phase factor is chosen so that it is equal to the classical action \cite{Kostant72preprint,Klein18qsmf,Bondar19prsa}
\begin{align} \label{eq:semiclassical_phaseaction}
\hbar\dot\phase=-\PhaseAction= \Lagrangian=\pcoord_j\dot\qcoord^j   - \Ham.
\end{align}
In this case, using integration by parts, the difference between $\pdf_\Ham$ and $\pdf$  is determined to be  \cite{Bondar19prsa}
\begin{align} \label{eq:pdf_Ham}
\pdf_\Ham-\pdf= \partial_j \left[\pdf J^{jk} (\hbar\partial_k\phase-\alpha_k)\right].
\end{align}
As proven in App. \ref{sec:HamiltonJacobi}, because the semi-classical phase factor also satisfies the Hamilton-Jacobi equation, 
%\begin{align}
%\left. \hbar \partial \phase/\partial{\qcoord^j} \right|_{q_0,\time}&=\pcoord_j &
%\left. \hbar \partial_\time\phase\right|_{q_0,q}& =-\Ham(\qcoord^j,\pcoord_j,\time),
%\end{align}
 the difference, $\pdf_\Ham-\pdf$, vanishes identically and  the final expressions for the   Hamiltonian in Eq.~\ref{eq:Wave_Hamiltonian_grad} and Eq.~\ref{eq:Wave_Hamiltonian_ddt}, precisely yield the classical energy.  
For example, in action-angle coordinates, where $\Ham_0(\action)$ alone, the action is simply
\begin{align} 
\hbar\phase = \action_j(\angle^j-\angle_0^j)-\Ham_0(\time-\time_0).
\end{align} 
Thus, the final expressions for the   Hamiltonian in Eq.~\ref{eq:Wave_Hamiltonian_grad} and Eq.~\ref{eq:Wave_Hamiltonian_ddt}, precisely yield the  classical energy density, $\Ham_0 \pdf$.  
This condition can be used to simplify a number of the results of Ref.~\cite{Bondar19prsa}.

\subsection{Symmetries and Conservation Laws}
Noether's theorem (explained in App. \ref{sec:symmetry}) states that a symmetry of the equations of motion leads to a conservation law for a corresponding density, $\CQ$.
For the KvN Lagrangian, there are a number of conservation laws in the  form
of a space-time divergence
\begin{align}
\partial_\time \CQ+\Vnabla\cdot \Vvel \CQ=0.
\end{align}
The invariance of the Hamiltonian under a constant change in phase factor, $\partial_\phase{\mathfrak H}$, implies that the KvN number density $\psi^\dagger\psi=\pdf$ is conserved. If the KvN Hamiltonian is invariant in time, $\partial_\time \Hhat$, then the KvN Hamiltonian density, $  \psi^\dagger \Hhat\psi$, is conserved. Similarly, if the KvN Hamiltonian is invariant with respect to coordinate, $\coord^j$, so that $\partial_j \Hhat=0$, then the KvN momentum density, $ \psi^\dagger \Phat_j \psi$, is conserved.

The conservation of probability density, $\pdf$, also implies that, if the system results from a classical Hamiltonian, $\Ham(\qcoord^j,\pcoord_j)$, that is independent of time, $\partial_\time \Ham=0$, then the classical energy density, $ \Ham \pdf$, is conserved. Similarly, if the classical Hamiltonian is independent of one of the coordinates, $\qcoord^j$, so that $\partial_{q^j} \Ham=0$, then the classical momentum density, $ \pcoord_j\pdf$ is conserved. 
  If $\PhaseAction$ is chosen to be a function of conserved quantities, e.g. $\Ham$ and/or $\pcoord_j$, etc.,  then the  classical Hamiltonian conservation laws imply that the  KvN counterparts also hold true.

\section{Quantum Simulation of Classical Dynamics \label{sec:QuantumSimulation}}
%\emph{4. Quantum Simulation of Classical Dynamics.---}
\subsection{Heisenberg Uncertainty }
%\emph{4. Heisenberg Uncertainty and Numerical Discretization.---}
The only nontrivial commutators in the extended phase space are 
\begin{align}
\Commutator{\hat\coord^j,\Phat_j} = i\hbar.
\end{align} 
Hence, the only nontrivial Heisenberg uncertainty relations are 
\begin{align}
  \sigma_{\coord^j}  \sigma_{\Mom_j}  \geq \hbar/2
\end{align}
where the uncertainty in quantity $A$ is defined via 
\begin{align}
\sigma_{A}^2 %= \avg{\left(\hat A^2 -\avg{\hat A}\right)^2}
 = \avg{\hat A^2}-\avg{\hat A}^2
.
\end{align}
 Thus, it is possible to make a simultaneous measurement of all of the classical phase space variables, $\coord^j$, or  all of the Lagrange multipliers, $\Mom_j$. 

\subsection{ Numerical Discretization }
{\it In this section, the Einstein summation convention is not used.} 

\begin{figure}
\includegraphics[width=3.375in]{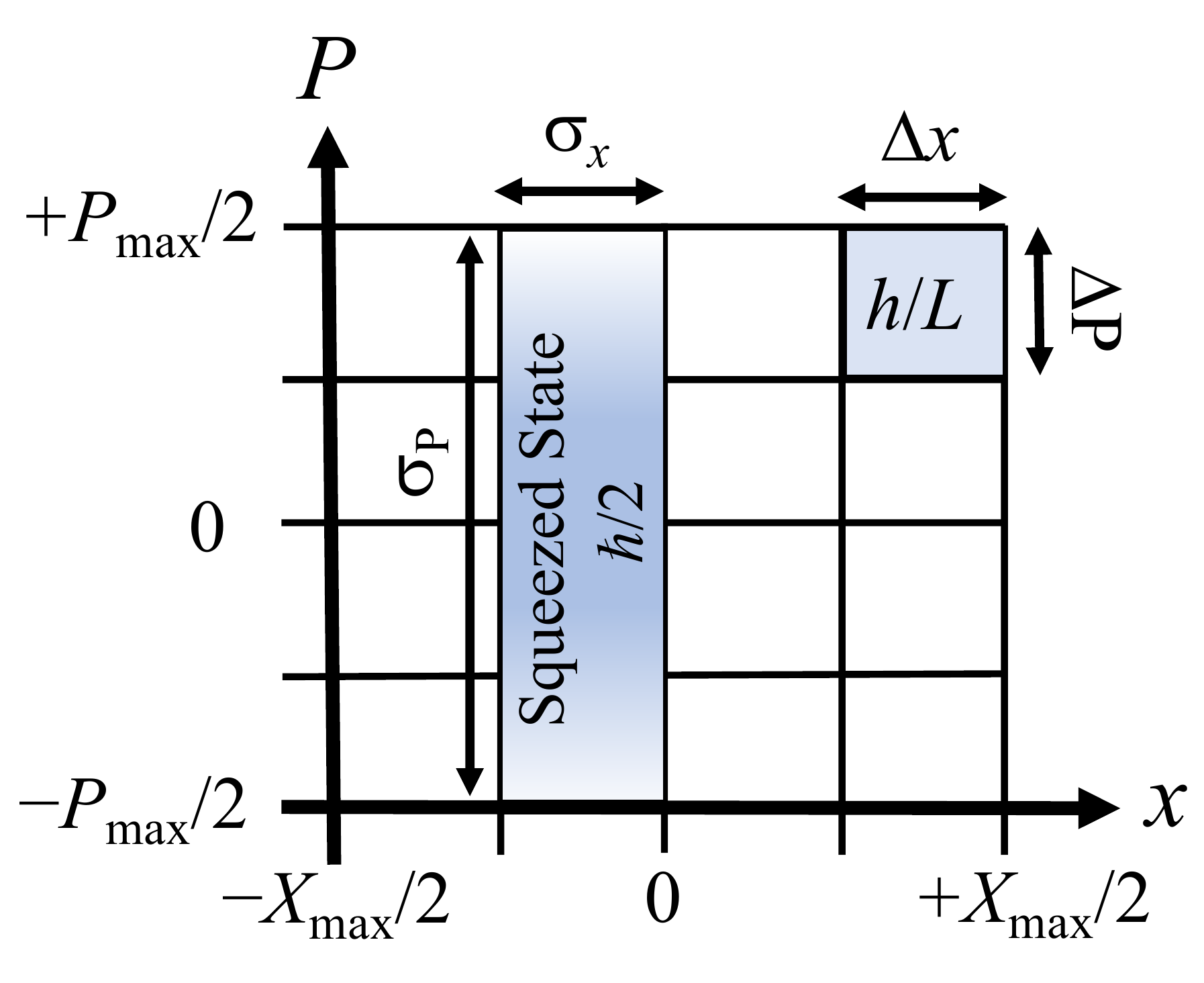}
\caption{A numerical discretization of phase space implies  finite numerical widths $\Delta\coord$ and $\Delta \Pcoord$.  However, the phase space area $\Delta\coord \Delta \Pcoord=\hPlanck/L$ due to the discretization is much smaller than than the Heisenberg limit allows.  Squeezed states can be used to reduce the  uncertainty to the numerical discretization limit, so that $\sigma_\coord=\Delta\coord$.}
\label{fig:numerical_discretization}
\end{figure}

In order to represent the classical phase space dynamics with a finite number of qubits, one must construct a finite-dimensional numerical approximation of the KvN Hamiltonian. 
 Assume that each coordinate, $\coord^j$, is periodic on the length,  $ \Xmax^j$, and is represented with an integer number of levels, $\Ngrid_j$, so that the level spacing is $\Delta \coord^j= \Xmax^j/\Ngrid_j$.  Then, the Fourier representation of the conjugate momenta, implies that these coordinates are also periodic and have $\Ngrid_j$ levels with level spacing $\Delta\Mom_j=\hPlanck / \Xmax^j$ and the range $\Mom_{j,{\rm max}}=\hPlanck \Ngrid_j/ \Xmax^j$. Thus, as illustrated in Fig. \ref{fig:numerical_discretization}, the phase space uncertainty due to the discreteness of the representation, $\Delta \coord^j\Delta \Mom_j  =  \hPlanck/\Ngrid_j$, is much  smaller than the Heisenberg limit allows.  

Often, one considers coherent states to be the analog of classical states. However, a coherent state, which saturates the Heisenberg bound, will have a width that is large compared to the classical level spacing, i.e. $\sigma_{\coord^j}/\Delta \coord^j = (\Ngrid_j/4\pi)^{1/2}$ and $\sigma_{\Mom_j}/\Delta \Mom_j = (\Ngrid_j/4\pi)^{1/2}$.  Since one is only interested in the dynamics of the original phase space, one can use squeezed states to reduce the uncertainty in the $\coord^j$ coordinates of interest and increase the uncertainty in the $\Mom_j$ coordinates. Squeezing the uncertainty in  $\coord^j$ by the factor of $\Ngrid_j^{-1/2}$ reduces the quantum uncertainty to the limit set by numerical discretization, $\sigma_{\coord^j}=\Delta \coord^j$. This increases the uncertainty in $\Mom_j$ by the factor $\Ngrid_j^{1/2}$ so that $\sigma_{\Mom_j} = \hbar \Ngrid_j / \Xmax^j=(\Ngrid_j/4\pi) \Delta\Mom_j$.
This implies that the relative uncertainty satisfies $\sigma_{\coord^j}/ \Xmax^j=1/\Ngrid_j$ while $\sigma_{\Mom_j}/\Mom_{j,{\rm max}}=1/4\pi$.
Thus, one can use squeezed states as initial conditions and as final \added{measurement} states in order to perform measurements that saturate the uncertainty limit set by numerical discretization.

\def\Ncopies{K}
%\emph{6. Quantum Simulation of Classical Dynamics.---}
\subsection{Complexity Estimates  \label{sec:Complexity}}
It is already known that there are quantum algorithms that can speed up the solution of a  linear system of ordinary differential equations \cite{Berry17arxiv} and linear partial differential equations, such as   wave equations \cite{Costa19arxiv}. Can this be extended to nonlinear systems?

\def\Nsteps{T}
\def\MCsparsity{r}
The KvN representation of classical dynamics implies that S. Lloyd's proof \cite{Lloyd96sci} that a quantum computer can be used to accelerate quantum simulation can also be applied to classical dynamics, even for non-Hamiltonian systems.  If the quantum Hamiltonian in Eq.~\ref{eq:KvN_Hamiltonian} is sufficiently sparse, e.g. because it is local, then one can break the operator into a small number, $m$, of non-commuting parts, $\Hhat=\sum_{j=1}^m \Hhat_j$, where $m$ is independent of \added{the number of states}, $\Ndof$. Application of the Trotter-Suzuki product formula can be used to generate an efficient simulation of the unitary evolution with error proportional to $\sum_{j\neq k} \avg{\Commutator{\Hhat_j,\Hhat_k}}$.  Recent results for general $s$-sparse Hamiltonian simulation \cite{Berry15focs} using $n$ qubits imply that the quantum simulation complexity\changed{, neglecting polylogarithmic factors,} is proportional to  $sn\added{T}$ \added{where $T=\norm{\norm{\Hhat \time}}_{\rm max}$ can be interpreted as the number of required time steps.} This implies large gains over an Eulerian discretization of the Liouville equation.

For example, consider the case of $\Nparticles$ particles that experience \added{$s$-sparse} local interactions between them, traveling in $2\xDim$ phase space dimensions, including both configuration space and momentum (or velocity) space, so that the total phase space dimension is $\Ndim=2\xDim \Nparticles$.  If  an Eulerian discretization is employed on a grid of $\Ngrid=2^\ell$ points in each direction, then this requires $\Ngrid^\Ndim=2^{\ell\Ndim}$ degrees of freedom, and $\added{\nqubit = } \ell\Ndim$ qubits to represent the PDF. If the kinetic energy is quadratic in velocity, then the maximum kinetic energy scales as $ \added{\Nsteps}\sim \xDim \Ngrid^2$, so that  
\begin{align} \label{eq:KvN_complexity}
\sparsity \nqubit \added{\Nsteps} \sim \sparsity \ell \Ndim \xDim \Ngrid^2 = 2\sparsity \ell \Nparticles  \xDim^2 \Ngrid^2.
\end{align}
 Thus, as compared to a  classical Eulerian discretization of the Liouville equation, this implies an exponential speedup in $\Ndim$ and a polynomial speedup in $\Ngrid$. If the number of dimensions, $\Ndim$,  is large, then  the degree of the polynomial speedup, $\Ndim/2$, is also large.  Similar   estimates can also be obtained for the max norm of the commutator.  
Note, however, that some important calculations can require a large number of time steps \cite{Benenti05book}, potentially scaling as a power of $\Ndim$, and this would reduce the expected savings to polynomial at best.  

However, the more interesting comparison is between the quantum simulation and the best probabilistic classical algorithm. For high dimensional PDEs, the best probabilistic classical algorithms that are known are typically some form of generalized time-dependent Monte Carlo (MC) algorithm, broadly including particle-based  techniques such as particle-in-cell (PIC) and molecular dynamics within the framework of Markov Chain Monte Carlo (MCMC). 
With appropriate assumptions, classical MC algorithms can also provide a large speedup over Eulerian discretization \cite{KalosMCbook, HeinrichNovak01arxiv} because the accuracy of the results will scale as  the square root of the number of samples, i.e. the computational work scales as $\sim \varepsilon^{-2}$, where $\varepsilon$ is the required accuracy. However, for quantum algorithms based on quantum walks, the accuracy generically decreases as the number of samples itself \cite{HeinrichNovak01arxiv, Aharonov02arxiv,Montanaro15prsa},  which leads to a quadratic improvement in computational work, $\sim\varepsilon^{-1}$, for a given accuracy.  

\added{
The computational complexity of an MC algorithm is similar in form to the estimate above for the quantum KvN algorithm. 
Since one must simulate $\Ncopies$ trajectories in $\Ndim=2\xDim \Nparticles$ dimensions for a number of time steps $\Nsteps$, with $r$-sparse interactions in the equations of motion per dimension, the complexity is 
\begin{align} \label{eq:MC_complexity}
 \Ncopies  \MCsparsity \Ndim \Nsteps   =    \Ncopies 2  \MCsparsity  \xDim  \Nparticles   \Nsteps.
\end{align} 
The key point is that the number of trajectories required to achieve a given accuracy $\varepsilon$ scales as $\Ncopies\sim \AsymptoticallyBoundedAboveBy(1/\varepsilon^2)$, so that the overall complexity is $\AsymptoticallyBoundedAboveBy(rDT/\varepsilon^2)$.
%Thus, it is  possible that   quantum simulation using the KvN approach to classical dynamics could lead to a quadratic improvement in computational cost relative to classical time-dependent MC algorithms.
}

\added{ 
In order to estimate the value of an observable with a given specified accuracy, $\varepsilon$, the quantum simulation of the KvN Hamiltonian must also be repeated multiple times.
If one simply averages the result of multiple trials, then the  number of trials required to achieve a given accuracy follows the   classical sampling law, $\AsymptoticallyBoundedAboveBy(1/\varepsilon^2)$, and the overall complexity would be similar to a classical MC algorithm \cite{Abrams99arxiv}. 
However, quantum algorithms based on amplitude amplification and estimation \cite{Grover98arxiv, Abrams99arxiv, Brassard00arxiv, Heinrich01jc, HeinrichNovak01arxiv, Brassard11arxiv, Montanaro15prsa} are able to compute numerical approximations to sums and integrals with a quadratic speedup relative to classical probabilistic algorithms, i.e. the accuracy generally decreases as the number of samples itself so that the number of times that the quantum simulation must be repeated typically follows the ``quantum sampling law,''  $\Ncopies \sim  \AsymptoticallyBoundedAboveBy(1/\epsilon)$, up to polylogarithmic factors. 
}

\def\RotatePDFOp{\hat R} 
\added{ 
The expectation value of an observable, $\avg{\Observable}$, is determined by the sum 
\begin{align}
\avg{\Observable}=\sum_x\Observable(x)\pdf(x).
\end{align} 
For any function on phase space, $\phi(x)$, define the state $\ket{\phi}$ via
\begin{align}
\ket{\phi}&= \sum_x\phi(x)\ket{x}/ \sqrt{\sum_x\norm{\phi}^2}
%\left(\sum_x\norm{\phi}^2\right)^{1/2}
\end{align}
where $\ket{x}$ is an element of the computational Hilbert space  of dimension $N$.
The KvN simulation computes the state $\ket{\psi}$ where $\psi=\pdf^{1/2} e^{i\phase}$.
In order to compute an observable,  $\avg{\Observable}$,   append an ancillary qubit to the Hilbert space and  compute the state \cite{Abrams99arxiv, Brassard11arxiv, Montanaro15prsa}
\begin{align}
 \RotatePDFOp_\phi  \ket{\amp}\ket{0} &= N^{-1/2}\sum_x \ket{x} \left( \phi' (x) \ket{0} + \phi(x) \ket{1}\right)
\end{align}
defined by
\begin{align}
	\phi(x)&:=  \Observable^{1/2}\psi \\
	\phi'(x)&:=\left(1-\abs{\phi}^2\right)^{1/2}e^{i\phase'}
\end{align}
Given $\ket{\psi}$, this state can readily be computed by using a quantum computer to simulate a reversible classical computation of $\phi$. 
The reversible calculation requires two KvN simulations: one to compute $\phi=\Observable^{1/2} \psi$ and one to uncompute $\phi$, which requires running the KvN simulation backward in time.
Amplitude estimation \cite{Brassard00arxiv} of the amplitude of the ancillary $\ket{1}$ state then yields an efficient estimate of the value of $\avg{\Observable}$.
This is  easily seen after rewriting the operation $\RotatePDFOp_\phi$ as a rotation of the target space
\begin{align}
 \RotatePDFOp_\phi 
\left(\begin{array}{l}
  \ket{\amp}\ket{0} \\
  \ket{\amp}\ket{1} 
\end{array}\right) 
 &=  
\left(\begin{array}{rr}
  \cos{(\theta)} & \sin{(\theta)} \\
  -\sin{(\theta)} & \cos{(\theta)}
\end{array} \right)
\left(\begin{array}{l}
  \ket{\phi'}\ket{0} \\
  \ket{\phi\,}\ket{1} 
\end{array}\right) 
\end{align}
by the angle $\sin{(\theta)}= \avg{\Observable/N}^{1/2}$.
Since each evaluation of $\RotatePDFOp_\phi$ and $\RotatePDFOp_\phi^\dagger$  uses two evaluations of $\ket{\psi}$, the amplitude amplification algorithm requires four KvN simulations to be performed per step.
Therefore, in order to achieve accuracy $\varepsilon$, the  KvN simulation must be repeated $4\Ncopies \sim \AsymptoticallyBoundedAboveBy(1/\epsilon)$ times and the overall complexity is $\AsymptoticallyBoundedAboveBy( s DT/\varepsilon)$. 
}

\added{
A number of useful quantum algorithms that can be directly applied to compute the value of an arbitrary bounded observable with a desired accuracy and a desired probability of success, as well as their complexity including polylogarithmic factors, are given by A. Montanaro in Ref. \cite{Montanaro15prsa}. In order to compute the value of an observable with either fixed absolute accuracy or fixed relative accuracy, one can use either Algorithm 3 or 4 of    \cite{Montanaro15prsa}, respectively. 
If the observable is non-negative, $\Observable(x)\geq 0$, such as an even moment of the PDF, then another choice is to apply Algorithm 2 of  \cite{Montanaro15prsa} directly. 
If the observable is bounded between 0 and 1, then one can use either Algorithm 1 of   \cite{Montanaro15prsa} or algorithms mean1 or mean2 of   \cite{Brassard11arxiv}.
The probability that a subset of phase space, $\CV$, is occupied  is an excellent example of the latter, because, in this case, the observable, $\Observable(x)$, is simply the indicator function that returns 1 if  $x\in \CV$ and 0 otherwise, and $\avg{\Observable}=\sum_{x\in\CV} \pdf(x)$.
}

\added{
Other algorithms in the literature can also be employed with minor modifications of the general procedure outlined above. For example,  the first algorithm  discussed in \cite{Abrams99arxiv} relies instead on performing a rotation by $\norm{\phi}^2=\Observable \pdf$. The second algorithm discussed in   \cite{Abrams99arxiv}, and used to prove Theorem 1 of \cite{Heinrich01jc}, uses the standard quantum counting algorithm \cite{Brassard00arxiv} by digitizing the computation of the value of $\Observable \pdf$ within an ancillary space of qubits.  As above, these algorithms can be performed by simulating a reversible classical computation of $\phi$.  
 }

%%%%%%

 In the typical case of quantum simulation and in the typical case of linear evolution, one can often argue that if the state of interest is smooth, e.g. because it is a ground state or a low-lying excited state, then the max norm may be a severe overestimate of the errors involved in computing the measured expectation values.  If the solution is well-resolved, then Fourier harmonics of the probability distribution should decay exponentially with increasing $\Ngrid$. This can potentially reduce the actual expectation value of the error to become exponentially small and can potentially imply an exponential speedup \cite{Parker20arxiv}. However, for nonlinear classical dynamics, the PDF generically develops  finely mixed phase space structure that approaches a singular and potentially fractal distribution in the infinite time limit \cite{LichtenbergBook}.  This implies that this kind of argument is not likely to apply in the setting of nonlinear classical dynamics.

There are also additional restrictions on the ability of quantum algorithms to speed up the calculation.  A general quantum simulation program must have  initialization, simulation, and  output (measurement) subprograms.  Care must be taken to ensure that the complexity of the initialization and output stages are less than or equal to the simulation stage.  This implies that it is not desirable to set each initial condition individually or to measure the entire PDF over all states.  The input states should be relatively easy to construct so that the initialization step is ``sparse.'' 
\added{
For example, it is often desirable initialize the PDF with a value that is close to an equilibrium, e.g. Maxwell-Boltzmann, distribution.  There are a number of algorithms based on quantum walks \cite{Szegedy04focs, Wocjan09pra, Montanaro15prsa} that approximate useful partition functions  with a quadratic speedup over classical MCMC algorithms.}
Similarly, the output should consist of a relatively small number of measurements, e.g. low-order moments of PDF, so that the measurement step is also ``sparse.''
\added{
As explained above, the measurement of a given physical observable as an average over phase space can also be performed with a quadratic speedup.  
Thus,  quantum simulation using the KvN approach to classical dynamics  leads to a quadratic improvement in computational cost relative to classical time-dependent MC algorithms.
}

 Finally, note that, if the underlying system is Hamiltonian and simulating the quantized Hamiltonian system is sufficient for the intended calculation, then, because the KvN approach leads to a system with twice the phase space dimension, simulating the quantized Hamiltonian   is the more efficient computational approach.   If the computational gains versus classical Monte Carlo are only quadratic, then doubling the phase space dimension would effectively eliminate the advantage. Moreover, while quantizing the Hamiltonian clearly yields self-consistent semiclassical dynamics,  the standard semiclassical approximation to the evolution operator \cite{KleinertBook,CvitanovicChaosBook} is not guaranteed to be unitary when multiple classical paths contribute to the result. Thus, a consistent approach to approximating the semiclassical evolution operator using the KvN framework has yet to be developed.

%\emph{7. Conclusion.---} 
\section{Conclusion \label{sec:Conclusion}}
In conclusion, quantum computers can be used to simulate nonlinear non-Hamiltonian classical dynamics on phase space by using the generalized Koopman-von Neumann  formulation of classical mechanics.  The Koopman-von Neumann formulation implies that the classical phase space dynamics expressed by the Liouville equation can be recast as an equivalent Schr\"odinger equation for the wavefunction on  Hilbert space.  The  wavefunction  completely specifies the probability distribution function and its dynamics is generated by a Hermitian Hamiltonian operator and a unitary evolution operator.  Thus, a quantum computer with finite resources can be used to simulate a finite-dimensional approximation of this unitary evolution operator. 

The conservation of probability on phase space can be expressed as an equivalent Schr\"odinger equation that is linear in momenta.  The equivalent Schr\"odinger equation corresponds to the quantization of a constrained Hamiltonian system with twice the dimension of the original phase space, where the conjugate momenta act as Lagrange multipliers that enforce the equations of motion on the original phase space.   Heisenberg's uncertainty principle applies between each variable and the corresponding Lagrange multiplier, but not between any of the variables of the original phase space. Squeezing the uncertainty of the original phase space variables and enhancing the uncertainty of the conjugate Lagrange multipliers allows the quantum uncertainty to be reduced to the limit set by numerical discretization.  Hence, there is complete fidelity to the classical phase space dynamics. 

Quantum simulation of the KvN representation of classical dynamics is exponentially more efficient than a \changed{deterministic} Eulerian discretization of the Liouville equation if the Koopman-von Neumann Hamiltonian is sparse.
\added{Quantum simulation of the KvN representation  is quadratically more efficient than a classical time-dependent Monte Carlo (MC) simulation. 
Many  useful initial states can  be prepared quadratically faster than classical MC by utilizing strategies based on quantum walks.
Utilizing quantum algorithms for computing sums and integrals leads to a quadratic speedup in the complexity of calculating  observables with a fixed accuracy.
Thus, up to polylogarithmic factors, the quantum simulation of the KvN Hamiltonian leads to an overall quadratic improvement in complexity relative to classical probabilistic algorithms.
}  
%It is possible that utilizing quantum walks and associated techniques could also lead to a quadratic improvement over classical time-dependent Monte Carlo methods.

Exploring the advantages and disadvantages of the Koopman-von Neumann representation for specific examples of classical dynamical systems and developing quantum simulation algorithms for    an accurate approximation to the semiclassical evolution operator are important directions for future work.

\section*{Acknowledgements} 
The author would like to thank J. L. DuBois, V. I. Geyko,  F. R. Graziani, S. B. Libby, J. B. Parker, \added{M. D. Porter}, N. A. Petersson, C. Tronci, Y. Shi, and K. Wendt for a number of interesting discussions on this subject. In particular, the author is grateful to N. A. Petersson and K. Wendt for carefully reading earlier versions of the manuscript and making suggestions that led to a number of improvements in the exposition, as well as to C. Tronci for sending B. Kostant's original 1972 conference paper on Prequantization. 
This work, LLNL-JRNL-807423, was performed under the auspices of the U.S. DOE by LLNL under Contract DE-AC52-07NA27344 and was supported by the DOE Office of Fusion Energy Sciences ``Quantum Leap for Fusion Energy Sciences'' project  FWP-SCW1680 and by LLNL Laboratory Directed Research and Development project 19-FS-078.

\appendix

\section{ Probability Distribution Function as a Volume Form \label{sec:PDF_volume_form}}

In the main body of the text, the probability distribution function (PDF), $\pdf(\Vx,\time)$, is defined as a volume form on the phase space  manifold, $\CM$ (see Refs.~\cite{SchutzBook, AbrahamMarsdenBook} for an introduction to the exterior calculus of differential forms on manifolds).
For any system of coordinates, $\Vx=\setlist{\coord^j}$, defined within a region of the manifold, the probability density measure is  defined by the   probability density   form  (PDF)
\begin{align}
\ProbabilityForm := \pdf \Vd\coord^1\wedge \Vd\coord^2\wedge \dots\wedge \Vd \coord^\xDim.
\end{align}
The probability satisfies the normalization condition that integration over the manifold, $\CM$, yields unit probability
\begin{align}
1 =\int_\CM \ProbabilityForm= \int_\CM \pdf d^\xDim\coord
\end{align}
where the final version sums over the coordinate charts of the manifold.
In any other coordinate system, $\Vy(\Vx,\time)$, the transformed PDF, $g(\Vy,\time), $ must satisfy 
\begin{align}
g(\Vy,\time)d^\xDim y =  \pdf(\Vx,\time) d^\xDim \coord.
\end{align}
Hence, the two PDFs are related by
\begin{align}
g(\Vy,\time)=  \abs{\Jac} \pdf(\Vx(\Vy,\time),\time)  ,
\end{align}
where $\Jac$ is the Jacobian, the determinant of the matrix of partial derivatives
\begin{align}
 {\Jac} = \det( \partial \coord^j/\partial \ycoord^k) .
\end{align}

The Liouville equation states that the PDF is an invariant volume form for the space-time velocity. In coordinates $X^\mu=(\time,\Vx)$, where the space-time velocity is $\Vcoord^\mu=(1,\Vvel)$,  the Liouville equation  can be expressed as
\begin{align}\label{eq:spacetime_liouville}
  \partial_\mu(\pdf V^\mu ):=   \partial_\time \pdf + \partial_j (\pdf \vel^j )=0
  .
\end{align}
This is proportional to the  space-time divergence of $\Vcoord^\mu$ with respect to the PDF, $\pdf$, defined by $\pdf^{-1} \partial_\mu(\pdf V^\mu )$. 

In order to simply the notation used in the main body of the text, for  any coordinate system, $\Vx=\setlist{\coord^j}$, the gradient operator, $\Vnabla$, is simply defined through the one-form of partial derivatives
\begin{align}
\Vnabla:= \Vd \coord^j  \partial_j
\end{align}
and the divergence symbol, 
$\Vnabla\cdot \Vvel$,  is simply defined   by  the expression
\begin{align}
\Vnabla\cdot \Vvel:=\partial_j\vel^j
.
\end{align}
The convention that yields the usual divergence operator appears when the PDF is treated as a scalar field, as discussed in App.~\ref{sec:PDF_scalar_form}.
 
 \section{ Probability Distribution Function as a Scalar Field  \label{sec:PDF_scalar_form}} 
When there is a natural phase space volume form, $\VolumeForm$, defined via
\begin{align}
 \VolumeForm :=  \Jac \Vd\coord^1\wedge \Vd\coord^2\wedge \dots\wedge \Vd \coord^\xDim,
\end{align}
then the volume density, $\Jac$,  accounts for coordinate transformations.  For example, if $\Jac_x$ is the volume density in $\Vx$ coordinates, then after transforming to  coordinates $\Vy(\Vx,\time)$, the volume density becomes $\Jac_y=\Jac_\coord\Jac^\coord_y$ where $ {\Jac}^\coord_y = \det( \partial \coord^j/\partial \ycoord^k)$ is the Jacobian of the transformation. 
 
 The Hilbert space inner product is now defined by the volume form
\begin{align}\label{eq:inner_product_scalar}
\left<\phase|\amp\right> := \int \phase^\dagger\amp  \VolumeForm = \int \phase^\dagger\amp \abs{\Jac} d^\xDim \coord
\end{align}
   In this case,   the probability density form can be defined as
  \begin{align} 
 \ProbabilityForm := \PDF\VolumeForm =\PDF \Jac \Vd\coord^1\wedge \Vd\coord^2\wedge \dots\wedge \Vd \coord^\xDim
\end{align} 
 where $\PDF$ is  a scalar function on phase space. The Liouville equation can then be written as 
\begin{align}
\partial_\mu( \pdf V^\mu ):= \partial_\time (\PDF \Jac  ) +  \partial_j ( \PDF \Jac  \vel^j) =0.
\end{align}

If the wavefunction, $\Amp$, is   defined as a scalar field via
\begin{align} \label{eq:scalar_wavefunction}
\Amp:=\PDF^{1/2}  e^{i\phase}
\end{align}
so that  $\PDF=\Amp^\dagger\Amp$, then it satisfies the KVN equation
\begin{align} \label{eq:scalar_KvN_equation}
\dot \Amp+\tfrac{1}{2} \Amp\Jac^{-1} \left( \partial_\time\Jac+ \partial_j  \Jac \vel^j\right) 
 =i \Amp  \PhaseAction/\hbar.
\end{align}

\def\Khat{{\hat\CK} }
Although  $\VPhat$ is no longer   Hermitian over the inner product given by Eq.~\ref{eq:inner_product_scalar}, a Hermitian momentum operator can still be defined as
\begin{align}\label{eq:Pihat}
\VLhat&:=  \Jac^{-1/2} \VPhat \Jac^{1/2} =  \Jac^{-1}\tfrac{1}{2}\left(\VPhat \Jac+\Jac\VPhat\right).
\end{align}
Similarly, the Hamiltonian operator that is Hermitian over the space-time measure, $\Jac d^\xDim\coord d\time$,  can be defined as
\begin{align}
  \Khat= \Jac^{-1/2} i\hbar \partial_\time \Jac^{1/2} = \Jac^{-1}\tfrac{1}{2}i\hbar \left(\partial_\time \Jac+\Jac\partial_\time \right).
\end{align}
Thus, the Hermitian Koopman-von Neumann Hamiltonian for $\Amp$ is
\begin{align} \label{eq:scalar_KvN_Hamiltonian}
\Khat
&=\tfrac{1}{2}\Jac^{-1}  \left(\VPhat\cdot \Jac \Vvhat +\Jac  \Vvhat\cdot \VPhat \right)   + \hat \PhaseAction \\
&=\tfrac{1}{2}  \left(\VLhat\cdot \Vvhat  +    \Vvhat\cdot \VLhat \right)   + \hat \PhaseAction
. 
\end{align}
This should be compared to Eq.~\ref{eq:KvN_Hamiltonian}.

Consider the   transformation from Cartesian coordinates to another coordinate system such as cylindrical or spherical coordinates.
Iff the Jacobian  is  independent of time, then one can divide  the  Liouville equation through by the Jacobian to find  
\begin{align}
  \partial_\time \PDF+ \Jac^{-1} \partial_j(\PDF \Jac  \vel^j) =0.
\end{align}
In this case, the equations in the main body of the text are also correct if one makes the replacement $\pdf\rightarrow \PDF$, $\psi\rightarrow \Psi$, and uses the usual definition of the  divergence symbol 
\begin{align}
\Vnabla_\Jac\cdot \Vvel:=  \Jac^{-1} \partial_{j} ( \Jac  \vel^j).
\end{align}
Note, however, that for time-dependent coordinate transformations, the Jacobian is generically time-dependent and one cannot divide through by the Jacobian.  

\section{Probability Distribution Function on Space-Time \label{sec:PDF_space_time}}
For a completely general space-time coordinate transformation from $X^\mu= \setlist{\time,\Vx}$ to $Y^\mu = \setlist{s(\Vx,\time),\Vy(\Vx,\time)}$ the   space-time velocity is  
\begin{align}
\Vcoord^\mu=\setlist{\Vcoord^0,\Vcoord^j}=\Vcoord^0\setlist{1,\Vvel} :=\setlist{\dot s,\dot \Vy}, 
\end{align}
the full space-time Jacobian is $\CJ = \det{(\partial X^\mu/\partial Y^\nu )}$, and  the space-time divergence is $\partial_\mu ( \Jac  \Vcoord^\mu)$.
 Thus, the  Liouville equation becomes
\begin{align}
\partial_\mu( \pdf \Vcoord^\mu ):= \partial_0 (\PDF \Jac  \Vcoord^0  ) +  \partial_j ( \PDF\Jac  \Vcoord^j) =0.
\end{align}

The PDF, $\pdf=\PDF\Jac$, is a $\xDim+1$-form that represents the probability distribution function over space and time. However, the conserved probability distribution function over phase space alone, $\pdf V^0=\PDF\Jac V^0$, is the spatial component of a $\xDim$-form.  If one defines the wavefunction via
\begin{align}
  \psi &=(\pdf V^0)^{1/2}e^{i\phase} &
  \Psi&=(\PDF V^0)^{1/2}e^{i\phase}
\end{align}
then one arrives at the KvN equations in Eq. \ref{eq:KvN_equation} and Eq. \ref{eq:scalar_KvN_equation}. The KvN Hamiltonian operators in Eq. \ref{eq:KvN_Hamiltonian} and Eq. \ref{eq:scalar_KvN_Hamiltonian} are Hermitian with respect to the corresponding Hilbert space inner products.

Alternatively, if one defines the wavefunction via
\begin{align}
  \psi &=\pdf^{1/2}e^{i\phase} 
  ,
\end{align}
then the KvN equation is
\begin{align}
\dot \amp+\tfrac{1}{2} \amp \partial_\mu   V^\mu&= \tfrac{1}{2}\left( \partial_\mu   V^\mu+V^\mu\partial_\mu \right)  \amp
 =i \amp  \PhaseAction/\hbar.
\end{align}
The Hamiltonian operator 
\begin{align}
\Hhat \psi &:=i\hbar \left[ \partial_0\psi +\tfrac{1}{2} (V^0)^{-1} \partial_0V^0\right]\\
&=-i\hbar  \left[ v^j\partial_j  +\tfrac{1}{2} (V^0)^{-1} \left( \partial_j V^j \right)\right]+\PhaseAction.
\end{align}
is Hermitian with respect to the Hilbert space inner product defined by
\begin{align}
 \left<\phase|\amp\right> &=\int \phase^\dagger\amp V^0 d^\xDim\coord.
\end{align}

If the wavefunction is defined as a space-time scalar
\begin{align}
  \Psi &=\PDF^{1/2}e^{i\phase} 
  ,
\end{align}
then the KvN equation is
\begin{multline}
\dot \Amp+\tfrac{1}{2} \Amp \CJ^{-1} \partial_\mu \CJ  V^\mu  =  \tfrac{1}{2} \CJ^{-1} \left( \partial_\mu  \CJ  V^\mu+\CJ V^\mu\partial_\mu \right)   \Amp
\\
 =-i    \Amp \PhaseAction/\hbar
 .
\end{multline}
The Hamiltonian operator 
\begin{align}
\Khat \Amp&:=i\hbar \left[\partial_0\Amp +\tfrac{1}{2}(\Jac V^0)^{-1}\partial_0(\Jac V^0)\right]
\\
&=-i\hbar  \left[ v^j\partial_j  +\tfrac{1}{2}(\Jac V^0)^{-1} \left( \partial_j \CJ V^j \right)\right]+\PhaseAction.
\end{align}
is Hermitian with respect to the Hilbert space inner product defined by
\begin{align}
 \left<\phase|\amp\right> &=\int \phase^\dagger\amp V^0 \Jac d^\xDim\coord.
\end{align}

\section{ Conservation of the Symplectic Form,  Poincar\'e Form, and Poisson Tensor \label{sec:symplectic_form_conservation}}

For generalized Hamiltonian and variational systems, the equations of motion, $\dot\Vx=\Vvel$,  defined implicitly via
\begin{align}
 \LagrangeForm\cdot \Vvel=\Vd\Ham+\partial_\time\Valpha
\end{align}
preserve the symplectic form
\begin{align}
d\LagrangeForm/d\time =0.
\end{align}
Using Cartan's formula for the action of the Lie derivative $\CL_{\Vvel}$ on the symplectic form yields
\begin{align}
d\LagrangeForm/d\time &=\partial_\time\LagrangeForm + \CL_\Vvel \LagrangeForm \\
&= \partial_\time\LagrangeForm  + \Vd\left(\Vvel\cdot \LagrangeForm  \right)+ \Vvel\cdot \Vd\LagrangeForm 
.
\end{align}
The final term vanishes due the fact that the symplectic form is closed, $\Vd\LagrangeForm=\Vd^2\Valpha=0$. The first two terms cancel due to the equations of motion.

The time derivative of the Poincar\'e form is
\begin{align}
d\PoincareForm/d\time=\partial_\time\PoincareForm+\Vvel \cdot\LagrangeForm+\Vd(\Vvel\cdot\PoincareForm).
\end{align}
Due to the equations of motion, the time derivative is the differential of the Lagrangian
\begin{align}
d\PoincareForm/d\time=\Vd(\Vvel\cdot\PoincareForm-\Ham)=\Vd\Lagrangian.
\end{align}
Since $\Vd^2\Lagrangian=0$, this yields an alternate derivation of the conservation of the symplectic form.

Conservation of the symplectic 2-form also implies conservation of the Poisson 2-vector, due to the definition, $\PoissonForm=\LagrangeForm^{-1}$.

\section{ Symplectic Volume Form for Generalized Hamiltonian and Variational Systems \label{sec:PDF_canonical_volume_form}}
The velocity defined by Hamilton's equations is divergence-free in canonical coordinates. For generalized Hamiltonian and variational equations of motion, the velocity is divergence-free with respect to a preferrred canonical volume form.

The  symplectic 2-form, $\LagrangeForm$, defines the canonical symplectic volume form via 
\begin{align}
 \VolumeForm := (\LagrangeForm\wedge)^\xDim = \Jac \Vd\coord^1\wedge \Vd\coord^2\wedge \dots\wedge \Vd \coord^\xDim.
\end{align}
Due to  the conservation of the symplectic 2-form derived in App.~\ref{sec:symplectic_form_conservation},
the symplectic volume form is also conserved.
This implies that symplectic volume density is  conserved by the flow
\begin{align}
d\Jac/d\time=\partial_\time \Jac + \Vnabla_\Jac\cdot \Vvel=\partial_\time \Jac+\partial_j (\Jac \vel^j)=0
\end{align}
and, hence, that the space-time divergence of the velocity with respect to the canonical volume form  vanishes.

\iffalse
Thus, let the Hilbert space inner product be defined by the canonical volume form
\begin{align}
\left<\phase|\amp\right> := \int \phase^\dagger\amp  \VolumeForm = \int \phase^\dagger\amp \abs{\Jac} d^\xDim \coord
\end{align}
and let  the canonical probability density form  
\begin{align}
\ProbabilityForm := \PDF  \VolumeForm = \PDF  \Jac   \Vd\coord^1\wedge \Vd\coord^2\wedge \dots\wedge \Vd \coord^\xDim.
\end{align}
be defined by the scalar field, $\PDF$. 
The wavefunction is also defined as a scalar field via 
\begin{align}
\Amp:=\PDF^{1/2}e^{\i\phase}
\end{align}
so that $\PDF=\Amp^\dagger\Amp$.
\fi
 
Thus, let the Hilbert space inner product be defined via Eq. \ref{eq:inner_product_scalar} so that  wavefunction, $\Amp$, defined in Eq. \ref{eq:scalar_wavefunction} and the PDF, $\PDF=\Amp^\dagger\Amp$, are treated as scalar fields.
 For generalized Hamiltonian mechanics, the Koopman-von Neumann Hamiltonian  for the scalar wavefunction $\Amp$ is
\begin{align}
\Hhat &=  \Jac^{-1} \VPhat\cdot   \Vvhat \Jac +\hat \PhaseAction  \\
&=\Jac^{-1/2}\VLhat\cdot \Vvhat\Jac^{1/2}+\hat \PhaseAction
\end{align}
and should be compared to Eq.~\ref{eq:hamiltonian_HKvN}. 
This Hamiltonian is Hermitian with respect to the canonical measure $\Jac d^\xDim\coord$ which is used in the definition of the inner product above.
Here, $\VLhat=\Jac^{-1/2}\VPhat\Jac^{1/2}$ is the Hermitian analog of the $\VPhat$ operator defined in Eq.~\ref{eq:Pihat}.

The Hermitian analog of the $\VXhat$ operator can be defined as 
\begin{align}
\boldsymbol \VZhat&= \tfrac{1}{2}\left(  \VLhat \cdot\hat \PoissonForm-\hat\PoissonForm\cdot\VLhat \right)\\
&=\tfrac{1}{2} \Jac^{-1/2}\left( \VPhat \cdot \hat \PoissonForm -\hat \PoissonForm\cdot\VPhat\right)\Jac^{1/2} \\
&=\tfrac{1}{2}\Jac^{-1} \left(\VPhat \cdot \Jac \hat \PoissonForm-\Jac \hat \PoissonForm\cdot \VPhat\right).
\end{align}
For general variational systems, the Koopman-von Neumann Hamiltonian  for the scalar wavefunction $\Amp$  is
\begin{align}
\Hhat &=  \Jac^{-1} \boldsymbol \VPhat\cdot \hat \LagrangeForm \cdot \Vvhat \Jac  +\hat \PhaseAction\\
&=\Jac^{-1/2} \boldsymbol \VZhat\cdot \hat \LagrangeForm \cdot \Vvhat \Jac^{1/2} +\hat \PhaseAction
\end{align}
and should be compared to Eq.~\ref{eq:variational_HKvN}.

Note, however, that the generalized form of Hamiltonian mechanics is often used in cases where the symplectic 2-form is degenerate \cite{Olver93book,Morrison98rmp}. In addition, while PDEs can be often be presented naturally in Hamiltonian form,  closed form expressions for the symplectic 2-form may either be unknown or rather difficult to construct. In such cases, the conventions used in the main body of the paper may be preferable.  

\section{ Symmetries and Noether's Theorem \label{sec:symmetry}}
The variational equations of motion can also be written as
\begin{align}
V\cdot dA=0
\end{align}
where they are generated by the space-time 1-form
\begin{align}
A=\Valpha-\Ham d\time=\alpha_jd\coord^j-\Ham d\time.
\end{align}

The derivation of Noether's theorem begins from the definition that a symmetry of the equations of motion is defined to leave $dA$ invariant. If the symmetry is generated by the space-time vector $U=\setlist{U^0,U^j}$, then   $\CL_UdA=0$  implies that, locally, 
\begin{align}
dS=\CL_UA=U\cdot dA+d(U\cdot A)
\end{align}  
for some scalar function $S$. Thus, taking the inner product with the solution to the equations of motion, $V$, yields the conservation law, 
\begin{align}
dQ/d\time=\CL_VQ=0,
\end{align}
 where
\begin{align}
 Q=V \cdot A - S.
\end{align}

\section{   Hamilton-Jacobi Equation \label{sec:HamiltonJacobi}}
In order to determine a complete solution to the Hamilton-Jacobi equation, begin with the solution to the equations of motion in action-angle coordinates $\setlist{\angle^j,\action^j}$. In action-angle coordinates, the Hamiltonian, $\Ham_0(\action)$, is a function of the action variables alone and the equations of motion
\begin{align}
\dot \angle^j=\omega^j_0&:=\partial_{\action^j}\Ham_0 & \dot\action^j=0
\end{align}
can be solved explicitly. In these coordinates, the action integral is simply
\begin{align}
\hbar  \phase&=\action_{j,0} (\theta^j-\theta^j_0) -  \Ham_0(\action_0) (\time-\time_0).
\end{align}
This satisfies the Hamilton-Jacobi equation and yields the canonical transformation between the action-angle coordinates and the initial conditions, $\setlist{\angle_0^j,\action_{j,0}}$.

Consider a canonical transformation to any other set of canonical coordinates, $\setlist{\qcoord^j,\pcoord_j}$, generated by the mixed variable generating function, $\GenFunc(\qcoord,\action,\time)$, that  is defined by the relations
\begin{align}
\left.  \partial \GenFunc/\partial{\qcoord^j} \right|_{\action,\time}&=\pcoord_j \\
\left. \partial \GenFunc/\partial{\action_j} \right|_{\time,\qcoord} &=\angle^j -\angle^j_0\\
\left. \partial \GenFunc/\partial \time \right|_{\qcoord,\action}&=\Ham_0(\action)-\Ham(\qcoord,\pcoord,\time)
.
\end{align}
This yields the action integral
\begin{align}
\hbar d \phase&=d\GenFunc-(\theta^j-\theta^j_0) d\action_j -  \Ham_0(\action) d\time
\\
&=d(\GenFunc-\Ham_0 \time) -(\theta^j-\theta^j_0-\omega^j_0 \time) d\action_j .
\end{align}
Due to the fact that the action coordinates, $\action_j$, are constants of the motion, this can be integrated to yield
\begin{align}
\hbar \phase=\hbar \phase_0+ \GenFunc(\qcoord,\action,\time)-  \Ham_0(\action)\time.
\end{align}
The choice $\phase_0=0$ implies that the action integral satisfies the Hamilton-Jacobi equation 
\begin{align}
\left. \hbar \partial \phase/\partial{\qcoord^j} \right|_{\action,\time}&=\pcoord_j\\
\left. \hbar \partial \phase/\partial \time \right|_{ \qcoord,\action }& =-\Ham(\qcoord ,\pcoord ,\time)
\end{align}
as well as the relation
\begin{align}
\left. \hbar \partial \phase/\partial{\action_j} \right|_{\time,\qcoord}&=\angle^j- \angle_0^j-\omega_0^j\time.
\end{align}
Hence, the partial derivatives, $\partial \phase/\partial{\action_j} =0$, vanish when evaluated along the trajectory, and, when evaluated along the  trajectory, the solution to the Hamilton-Jacobi equation also satisfies the partial differential equations
\begin{align}
\left. \hbar \partial \phase/\partial{\qcoord^j} \right|_{\pcoord,\time}&=\pcoord_j \\
\left. \hbar \partial \phase/\partial \time \right|_{ \qcoord,\pcoord }& =-\Ham(\qcoord ,\pcoord ,\time).
\end{align}

Thus, the choice $\phase_0=0$ implies that the final expressions for the   Hamiltonian in Eq.~\ref{eq:Wave_Hamiltonian_grad} and Eq.~\ref{eq:Wave_Hamiltonian_ddt}, precisely yield the  classical energy density, $\Ham \pdf$. This can be used to simplify a number of the results of Ref. \cite{Bondar19prsa}.

One can also prove that, if $\pdf_\Ham-\pdf$,  as given by Eq. \ref{eq:pdf_Ham}, vanishes as an initial condition, then it vanishes for all time.
Equation   \ref{eq:pdf_Ham} can be written as the constraint
\begin{align}
C=\Vd \pdf\cdot \PoissonForm\cdot (\hbar \Vd\phase-\Valpha) =0 .
\end{align}
The fact that, for any scalar, $S$, 
\begin{align}
\CL_V dS= d(V\cdot d S).
\end{align}
 implies that $\CL_Vd\pdf=0$. Combining this with the relations, $d\PoissonForm/d\time=\CL_V\PoissonForm=0$ and  $d\Valpha/d\time=\CL_V\Valpha=\Vd\Lagrangian$, proven in App. \ref{sec:symplectic_form_conservation} above, yields 
\begin{align}
dC/d\time=\CL_VC = \Vd \pdf\cdot \PoissonForm\cdot\left( \hbar \Vd\dot\phase-\Vd\Lagrangian\right).
\end{align}
This  vanishes due to the identification of the semiclassical phase factor with the classical action,  $\hbar\dot\phase=\Lagrangian$.

% .bib files must eventually be converted to .bbl files for submission

%\bibliography{divrmp_pop}	
%FOR NOW USE
%\include{Quantum-Classical-Bibliography.tex}
\bibliography{Quantum-Classical-Bibliography_v3.bib}{}
\vfil
\eject
\end{document}